\documentclass[11pt]{article}

\usepackage[pdftex]{graphicx}
\usepackage{amsfonts}
\usepackage{setspace}
\usepackage{amssymb}
\usepackage{amsthm}
\usepackage{graphicx}
\usepackage{amsmath}
\usepackage{mathtools}
\usepackage{lscape, lineno}
\usepackage{subcaption}
\usepackage{suffix}
\usepackage{color}
\usepackage{bm}
\usepackage{verbatim}
\usepackage{tensor}

\usepackage{enumitem} 
\usepackage{soul}

\usepackage{xr} %% for cross-reference 
\newcommand{\blind}{1}

\definecolor{darkblue}{rgb}{0.0,0.0,0.55}
\RequirePackage[colorlinks,citecolor=darkblue,urlcolor=darkblue,linkcolor=darkblue]{hyperref} 
\usepackage[normalem]{ulem}

\newtheoremstyle{compactplain}
  {2pt}{2pt}{\itshape}{}{\bfseries}{.}{5pt plus 1pt minus 1pt}{}
\newtheoremstyle{compactremark}
  {2pt}{2pt}{\normalfont}{}{\itshape}{.}{5pt plus 1pt minus 1pt}{}
\theoremstyle{compactplain}

\newtheorem{theorem}{\textsc{{Theorem}}}%[section]
\newtheorem{example}{\textsc{{Example}}}%[section]
\newtheorem{proposition}{\textsc{{Proposition}}}%[section]
\theoremstyle{compactremark}
\newtheorem{remark}{{\bf \textsc{Remark}}}%[section]

\usepackage{booktabs}
\usepackage{array}
\usepackage{multirow}
\newcommand{\bfe}{\mathbf{e}}
\newcommand{\bfh}{\mathbf{h}}

\newcommand{\bfm}{\mathbf{m}}

\newcommand{\bfs}{\mathbf{s}}

\newcommand{\bfv}{\mathbf{v}}

\newcommand{\bfx}{\mathbf{x}}

\newcommand{\bfA}{\mathbf{A}}

\newcommand{\bfD}{\mathbf{D}}

\newcommand{\bfI}{\mathbf{I}}

\newcommand{\bfK}{\mathbf{K}}

\newcommand{\bfO}{\mathbf{O}}

\newcommand{\bfQ}{\mathbf{Q}}

\newcommand{\bfV}{\mathbf{V}}

\newcommand{\bfY}{\mathbf{Y}}

\newcommand{\Ell}{\boldsymbol \ell}

\newcommand{\bfeta}{\boldsymbol \eta}
\newcommand{\bfxi}{\boldsymbol \xi}

\newcommand{\bfvartheta}{\boldsymbol \vartheta}

\newcommand{\bflambda}{\boldsymbol \lambda}
\newcommand{\bfmu}{\boldsymbol \mu}

\newcommand{\bfzeta}{\boldsymbol \zeta}

\newcommand{\bfomega}{\boldsymbol \omega}

\newcommand{\bfSigma}{\boldsymbol \Sigma}
\newcommand{\bfLambda}{\boldsymbol \Lambda}

\newcommand{\diag}{\text{diag}}

\newcommand{\calM}{{\cal M}}
\newcommand{\calK}{{\cal K}}

\newcommand{\calU}{{\cal U}}

\usepackage[sort,comma]{natbib}

\setcitestyle{citesep={;}}

\renewcommand{\baselinestretch}{1.5}

\allowdisplaybreaks

\def\spacingset#1{\renewcommand{\baselinestretch}%
{#1}\small\normalsize} \spacingset{1}

\expandafter\def\expandafter\normalsize\expandafter{%
    \normalsize%
    \setlength\abovedisplayskip{2pt}% reduce the spacing before equation
    \setlength\belowdisplayskip{2pt}% reduce the spacing after equation
    \setlength\abovedisplayshortskip{-8pt}%
    \setlength\belowdisplayshortskip{2pt}%
}

\usepackage{titlesec}
\titlespacing*{\section}{0pt}{10pt}{0pt} % Adjust values as needed
\titlespacing*{\subsection}{0pt}{0pt}{0pt} % Adjust values as needed

\titleformat{\section}
  {\normalfont\fontsize{14pt}{16pt}\bfseries} % Format for the entire line (size, font weight)
  {\thesection}                               % How to display the section number
  {1em}                                       % Space between the number and the title text
  {}                                          % Formatting for just the title text (can be left empty)
  []  
  
\titleformat{\subsection}
  {\normalfont\fontsize{14pt}{16pt}\bfseries} % Format for the entire line (size, font weight)
  {\thesubsection}                               % How to display the section number
  {1em}                                       % Space between the number and the title text
  {}                                          % Formatting for just the title text (can be left empty)
  []  
  
  \usepackage{needspace}
  
\usepackage{float}

\begin{document}

\if1\blind
{
  \title{\bf Asymmetric Space–Time Covariance Functions via Hierarchical Mixtures}
  \author{Pulong Ma \\
Department of Statistics, Iowa State University\\
2348 Osborn Dr, Ames, IA 50011\\
 % plma@iastate.edu
}
      \date{}
  \maketitle

} \fi

\if0\blind
{
  \bigskip
  \bigskip
  \bigskip
  \begin{center}
    {\LARGE\bf Asymmetric Space–Time Covariance Functions via Hierarchical Mixtures}
\end{center}
  \medskip
} \fi

\begin{abstract}
This work is focused on constructing space-time covariance functions through a hierarchical mixture approach that can serve as building blocks for capturing complex dependency structures.  This approach provides a unifying modeling framework that not only constructs a new class of asymmetric space-time covariance functions with closed-form expressions, but also provides corresponding space-time process representations, which further unify constructions for several existing space-time covariance models. This hierarchical mixture framework decomposes the complexity of model specification at different levels of hierarchy, for which parsimonious covariance models can be specified with simple mixing measures to yield flexible properties and closed forms. A characterization theorem is provided to study how the mixing measures determine the statistical properties of covariance functions. Several new covariance models resulting from this framework are discussed in terms of their practical usefulness. This hierarchical mixture approach is extended for nonstationary and multivariate settings. A theorem is also provided to construct valid asymmetric space-time covariance functions with arbitrary and possibly different degrees of smoothness and long-range dependence in space and in time. This generalized class also bridges a theoretical gap in using the Lagrangian reference framework. The superior performance of several new parsimonious covariance models over existing models is verified with the well-known Irish wind data and the U.S. air temperature data.
\end{abstract}

\newpage
\spacingset{1.65} % DON'T change the spacing!
%\doublespacing
%\onehalfspacing

\section{Introduction}\label{sec:intro}
Many geophysical processes often evolve in space over time, resulting in complicated space-time data with complex space-time dependence structures. For example, many atmospheric and oceanic processes move along a flow direction and hillslope creep exhibits long-memory dependence. Statistical analysis for such space-time data is often challenging in capturing such complex dependence and dealing with high-volume datasets; see \cite{Cressie2011} for an introduction to this topic from a hierarchical modeling perspective.    

Consider a real-valued stationary space-time process $\{y(\bfs,t): \bfs\in \mathbb{R}^d, t\in \mathbb{R}\}$. Let $K(\cdot,\cdot)$ be a positive-definite function such that $cov(y(\bfs_1,t_1 ),y(\bfs_2,t_2 ))=K(\bfs_1-\bfs_2,t_1-t_2)$ for all $\bfs_1,\bfs_2\in \mathbb{R}^d, t_1,t_2\in \mathbb{R}$. Then $K$ is called \textit{stationary}. If $K(\bfs_1-\bfs_2,t_1-t_2)$ only depends on the distance $\|\bfs_1 - \bfs_2\|$ in space, $K$ is called spatially isotropic. \cite{Gneiting2002} calls a stationary space-time covariance function $K(\mathbf{h},t)$ \textit{fully symmetric} if $K(\bfh,t)=K(\bfh,-t)=K(-\bfh,t)=K(-\bfh,-t)$. A space-time covariance function that is not fully symmetric (or simply asymmetric) is physically more realistic because of a dominant flow direction over time for modeling many geophysical processes. Although the restriction to stationary space-time covariance models is unrealistic for real-world applications, there are many different ways \citep{Fuentes2001, Higdon1998, Nott2002, Sampson1992, Stein2005nonstat, Paciorek2006} to build nonstationary space-time covariance functions. The main focus of this paper is to address the fundamental issue of constructing stationary covariance models with space-time interaction, since they can serve as building blocks for generating more complex and realistic space-time covariance models using various existing techniques. 

Earlier work on modeling space-time processes focuses on separable space-time models, where the space-time covariance function is a product of spatial covariance and temporal covariance, which may not be physically realistic for many geophysical processes. A further implication is that there is a source of discontinuity due to a lack of smoothness away from the origin of separable covariance functions than at the origin \citep{Stein2005}. Substantial efforts thus have been made in constructing nonseparable space-time covariance models  with a rich literature \citep{Cressie1999, Fonseca2011, Fuentes2008, Gneiting2002, Horrell2017, DeIaco2002, Jones1997, Ma2003JSPI, Ma2002Geo, DeLuna2005, Porcu2008, Stein2005, Stein2005JRSSB, Salvana2021, Porcu2020}; see \cite{Chen2021} and \cite{Porcu2020} for recent overviews. 

There are four main categories for constructing nonseparable space-time covariance models. The first category is the half-spectral method which provides great flexibility in constructing new classes of covariance models \citep[e.g.,][]{Cressie1999, Horrell2017, Stein2005JRSSB, Stein2005} by specifying different half-spectra forms that can allow possibly flexible behaviors such as smoothness near the origin and asymmetry.  The second category is to use completely monotone functions for constructing nonseparable space-time covariance models \citep{Gneiting2002}, but the resulting covariance models are fully symmetric and have a source of discontinuity \citep{Stein2005} as in the separable space-time covariance models. The third category is based on stochastic partial differential equations (SPDE). Based on the SPDE representation of the Mat\'ern covariance \citep{Whittle1963} and its inferential strategy via Gaussian Markov random fields \citep{Lindgren2011}, there are various constructions for space-time processes \citep[e.g.,][]{Sigrist2015, Cameletti2013, Lindgren2023}.  The fourth category is the mixture modeling approach. This approach was discussed by \citet[][pp.~10-37]{Matern1986} and more recent work \citep{Porcu2013, Ma2023,Schlather2010, Gneiting1997, Zhang2025} for constructing valid spatial covariance functions. This approach has been extended to produce space-time covariance models by taking mixtures of purely spatial covariances and purely temporal covariances \citep[e.g.,][]{Ma2002Geo, Ma2003JSPI, DeIaco2002, Porcu2008, Porcu2007, Fonseca2011}.

A nonseparable space-time covariance function does not necessarily imply asymmetry. 
The space-time covariance models via the mixture approach are often nonseparable; however, the resulting covariance models are fully symmetric and often lack flexible smoothness behaviors near the origin or long-range dependence. Approaches to induce space-time asymmetry include half-spectral methods \citep[e.g.,][]{Stein2005, Stein2005JRSSB, Horrell2017}, physical PDEs \citep{Kolovos2004, Jun2007, Jones1997, Brown2000}, and the Lagrangian reference framework \citep{Cox1988, Gupta1987}. The Lagrangian reference framework constructs a space-time process by moving a \textit{base} spatial process according to a velocity vector across time and satisfies Taylor's hypothesis \citetext{\citealp[p.~478]{Taylor1938}, \citealp{Gneiting2007chapter}} that has found wide interest in fluid dynamics, meteorology and hydrology. If this velocity vector is constant, the resulting space-time process is called a frozen field \citep[][p.~320]{Cressie2011}. A more flexible and realistic approach is to model the velocity vector probabilistically via a multivariate distribution. While this strategy is appealing, a key challenge is to derive \textit{closed-form} (or analytic) expressions via the multivariate distribution of the velocity vector in order to allow its wide application for statistical inference. One successful example is given by \cite{Schlather2010} with a closed-form expression based on a squared-exponential (or Gaussian) correlation function; however, a fundamental limitation in this space-time covariance function is that it is infinitely differentiable and decays exponentially in time. This sacrifices many desirable properties such as smoothness and long-range dependence for modeling space-time processes \citep{Stein2005}. It is also unclear how the resulting space-time covariance function is related to the original base spatial process in the Lagrangian reference framework, thus making it less interpretable. These issues will be addressed from a hierarchical modeling perspective to allow for simpler and more interpretable   construction. 

Exploiting the fundamental idea of hierarchical modeling \citep{Bernardo1994,Berger2013} for spatio-temporal data \citep{Cressie2011, Banerjee2014},  this paper presents a hierarchical mixture modeling framework that not only constructs a new class of asymmetric space-time covariance functions with {closed-form} expressions, but also provides corresponding space-time process representations, through which the complexity of the model specification can be decomposed at different levels of hierarchy. Here the closed form is in the sense that special mathematical functions could be used such as in the well-known Mat\'ern class \citep{Matern1986, Stein1999} since fast numerical algorithms for evaluation are often readily available in existing software. The key idea in the proposed hierarchical mixture framework is to mix space-time processes with mixing measures at different hierarchical levels, where the space-time process can be specified with relatively simple structures at each level to allow flexible modeling and closed-form expressions of covariance functions.

The proposed hierarchical mixture modeling approach is built upon a location and scale mixture technique - a well-known probabilistic strategy for generating new distributions \citep{Good1995,West1987, Barndorff1982, Gneiting1997} and covariance models \citetext{\citealp[pp.~10-37]{Matern1986}, \citealp{Ma2023,Schlather2010, Ma2002Geo, Ma2003JSPI, Porcu2007, Porcu2020, Porcu2013, FONSECA2011biometrika, Tang2024}}. The location mixture induces asymmetry between space and time and includes the Lagrangian reference framework as a special case; the scale mixture can allow flexible properties of the covariance functions such as smoothness behavior and long-range dependence. Starting with a base covariance function (or corresponding process representation) as in the Lagrangian framework, this hierarchical mixture approach then uses location mixture to induce asymmetry and scale mixtures to induce desirable statistical properties at different hierarchical levels, so that each level can focus on simple construction, which not only reduces the difficulty of analytical computation but also allows the flexibility in incorporating different mixing probability measures for desirable statistical properties. This is the essential idea in hierarchical modeling that is manifested in the proposed hierarchical mixture approach. As will be shown, a new class of asymmetric covariance models can be constructed with closed-form expressions and desirable properties, by designing mixing measures at different levels of hierarchy. This hierarchical mixture modeling approach provides a general and unified construction for several existing classes of space-time covariance models. The choice of mixing measures balances the simplicity of analytic derivation and the flexibility of statistical properties in the resulting covariance functions. A byproduct of this approach is that several specific space-time covariance models are derived with closed forms and a theoretical result that bridges a theoretical gap in \cite{Salvana2023} and \cite{Zhang2024}, who assumed the existence of closed forms and the positive-definite covariance functions via the Lagrangian  framework for arbitrary choice of base covariance functions. This paper also gives a characterization theorem to design mixing measures so that the resulting covariance functions are asymmetric with flexible origin and tail behaviors. This theorem offers insights on the choice of mixing measures that could allow flexible smoothness behavior near the origin like the Mat\'ern class  and/or long-range dependence in space and in time. 

The remainder of the paper is organized as follows. Section~\ref{sec: scale mixtures} presents a general hierarchical mixture modeling framework for generating new space-time covariance models with corresponding hierarchical construction for space-time processes. This framework is  also illustrated for several existing nonseparable space-time models by designing corresponding mixing measures. Theoretical properties of the space-time covariance models via hierarchical mixtures are further investigated in Section~\ref{sec: property} with recommendations for modeling space-time processes. Built on hierarchical mixtures, Section~\ref{sec: stcov} presents a new class of asymmetric space-time covariance models with flexible smoothness behavior and long-range dependence, and Section~\ref{sec: hierarchical mixtures of existing models} gives several examples of existing space-time covariance functions.  Numerical studies in Section~\ref{sec: numerical example} for the classic Irish wind speed data and the U.S. air temperature data confirm the advantages of the proposed new covariance models over existing models. Section~\ref{sec: discussion} concludes the paper with potential  improvements and research directions.

\section{A Hierarchical Mixture Modeling Framework} \label{sec: scale mixtures}

The proposed hierarchical mixture modeling framework is presented from two perspectives: one is based on space-time process representation, and the other is based on covariance function representation. 

\subsection{Space-Time Process Representation}
Let $\epsilon(\cdot)$ denote a spatial white-noise process on $\mathbb{R}^d$ with $\text{cov}(\epsilon(\bfs_1), \epsilon(\bfs_2))=\mathbb{I}(\bfs_1=\bfs_2)$ for any $\bfs_1,\bfs_2\in \mathbb{R}^d$, where $\mathbb{I}(\cdot)$ denotes an indicator function. Let $w(\cdot)$ denote a spatial process on $\mathbb{R}^d$ with a stationary covariance function $C_s(\cdot; \bfvartheta)$, where $\bfvartheta$ denotes its covariance parameters. Further assume that $w(\cdot)$ and $\epsilon(\cdot)$ are independent. Then we define the following space-time process $y(\cdot, \cdot)$:
\begin{align} \label{eqn: process rep}
y(\bfs, t) = \int_{\mathbb{R}^d} w(\bfs- \bfv t) \epsilon(\bfv) f(\bfv|\bfxi)\, \mathrm{d} \bfv, \quad (\bfs, t)  \in \mathbb{R}^d \times \mathbb{R},
\end{align}
where $f(\bfv|\bfxi)$ is a  kernel function in $\mathbb{R}^d$ depending on parameters $\bfxi$. It satisfies that (i) $f$ is a continuous function of $\bfv$ in $\mathbb{R}^d$, (ii) $f(\bfv|\bfxi)\geq 0$ for any $\bfv, \bfxi$, and (iii) $\int f(\bfv|\bfxi) \mathrm{d} \bfv<\infty$ for any $\bfxi$.  $y(\bfs, t)$ implicitly depends on parameters $\bfvartheta$ and $\bfxi$. The stochastic integral is defined in the sense of a mean-square limit.  To avoid confusion in the integral representation, $y(\cdot, \cdot)$ is also written as $y(\cdot, \cdot; \bfvartheta, \bfxi)$. The spatial process $w(\cdot)$ becomes a space-time process $\{w(\bfs-\bfv t): (\bfs, t) \in \mathbb{R}^d \times \mathbb{R}\}$ that moves along the direction of $\bfv$ and satisfies \textit{Taylor's hypothesis} \citep[][pp.~319-320]{Cressie2011}. $\{w(\bfs-\bfv t): (\bfs, t) \in \mathbb{R}^d \times \mathbb{R}\}$ represents a frozen field for any given $\bfv \in \mathbb{R}^d$. A key motivation in the space-time construction in~\eqref{eqn: process rep} is that it can be linked to certain physical partial differential equations (PDE); see \cite{Kolovos2004} for various examples.    

The representation~\eqref{eqn: process rep} has a close connection to existing spatial modeling frameworks with extensions to space-time settings: first, the process $y(\cdot, \cdot)$ can be viewed as a generalized process-convolution representation that extends several classes of existing process-convolution models \citep{Higdon1998, Higdon2003, Paciorek2006, Fuentes2001, Zhu2010} in spatial settings. The process $y(\cdot, \cdot)$ serves as a moving average process over the product of the space-time process $w(\bfs-\bfv t)$ and spatial white-noise process $\epsilon(\cdot)$ with the weights given by the kernel $f$. For each fixed $t$, the process $y(\cdot, \cdot)$ becomes a (spatial) process convolution representation. The choice of kernel $f$ can lead to very flexible class of covariance function models; see \cite{Fuentes2001} for an illustration in spatial settings. While the representation~\eqref{eqn: process rep} is focused on continuous convolution, discrete version can also be developed as well; see \cite{Fuentes2001} for an example.  Second, if one normalizes $f$ to a density, then the representation~\eqref{eqn: process rep} for $y(\cdot, \cdot)$ has a location mixture form, which is a key technique to induce asymmetry in the Lagrangian reference framework. This location mixture is equivalent to phase shift in the spectral representation of the space-time process; see \cite{Stein2005JRSSB} for an example. 

The third connection is that it generalizes the Lagrangian reference framework established by \cite{Cox1988} to construct asymmetric space-time covariance models. To see this connection, we first write out the covariance function of $y(\cdot, \cdot)$  in~\eqref{eqn: process rep}: 
\begin{align} \label{eqn: cov rep}
    C(\bfh, u; \bfzeta) :=\text{cov}\{y(\bfs_1, t_1), y(\bfs_2, t_2)| \bfxi, \bfvartheta \}    
    = \int_{\mathbb{R}^d} C_s(\bfh - \bfv u) f^2(\bfv|\bfxi)   \mathrm{d} \bfv.
\end{align}
where $\bfh = \bfs_1 - \bfs_2$ and $u = t_1 - t_2$ for $\bfs_1, \bfs_2 \in \mathbb{R}^d, t_1, t_2 \in \mathbb{R}$. $\bfzeta:=\{\bfxi, \bfvartheta\}$ denotes the covariance parameters. For notational convenience, $C(\mathbf{h}, u; \bfzeta)$ is also written as $C(\mathbf{h}, u)$ with the parameter dependence suppressed whenever there is no scope for confusion. This covariance function allows asymmetry with its parameters varying across space and time. To see its connection to the Lagrangian reference framework, let us consider a common construction: the Lagrangian reference framework defines a space-time covariance $\tilde{C}(\cdot, \cdot)$ by averaging the covariance $C_s(\mathbf{h} - \mathbf{v} u)$ with respect to a random velocity vector $\bfv$, that is, $\tilde{C}(\bfh, u):=E_{\mathbf{v}}[C_s(\mathbf{h} - \mathbf{v} u)]$. To match the form~\eqref{eqn: cov rep}, the probability density function of $\mathbf{v}$ can take the form of 
$\tilde{f}(\bfv|\bfxi) : = {f^2(\bfv|\bfxi)}/{\int_{\mathbb{R}^d} f^2(\bfx|\bfxi) \mathrm{d} \bfx}.
$
Then $\tilde{f}$ is always a valid density as long as the denominator is well defined. The covariance $\tilde{C}$ thus only differs from $C$ by a normalizing constant that depends on $\bfxi$. The covariance  $C$ thus can be viewed as a location mixture. This also gives the exact correspondence between the kernel $f$ in the process representation~\eqref{eqn: cov rep} and the probability density function $\tilde{f}$ in the Lagrangian reference framework. 

In the location mixture in~\eqref{eqn: process rep} or~\eqref{eqn: cov rep}, if the mixing distribution on the velocity vector $\bfv$ varies in space and in time, it can induce nonstationarity in a similar way as in \cite{Paciorek2006}, \cite{Stein2005nonstat} and \cite{Zhu2010}. The smoothness behavior of the process $y(\cdot, \cdot)$ depends crucially on the choice of mixing distribution as we show later. In fact, other statistical properties such as long-range dependence can also be achieved via a scale mixture over the parameters of $y(\cdot, \cdot)$. For instance, \cite{FONSECA2011biometrika} use scale mixture to obtain a nonstationary covariance for modeling non-Gaussian spatial data. From a hierarchical modeling perspective, we can design an additional hierarchical level for model specification by taking a scale mixture to allow for flexible tail behaviors such as long-range dependence.  

In particular, let $\bfeta$ denote a subset of $p$ parameters in $\bfzeta$ (such as the range parameter in $C_s$) via some probability measure $G$. Let $\delta(\cdot)$ denote a white-noise process in $\mathbb{R}^p$ with unit variance. Let $g(\bfeta|\bflambda)$ denote a kernel function in $\mathbb{R}^p$ that  depends on parameters $\bflambda$. Then applying the same idea in~\eqref{eqn: process rep} yields  a new space-time process $z(\cdot, \cdot)$ at another hierarchical level: 
\begin{align} \label{eqn: process rep z}
    z(\bfs, t) = \int_{\mathbb{R}^p} y(\bfs, t; \bfzeta) \delta(\bfeta) g(\bfeta| \bflambda)  \mathrm{d} \bfeta.
\end{align}
The covariance of $z(\cdot, \cdot)$ is thus given by 
\begin{align} \label{eqn: cov rep z}
{K}(\bfh, u) := \text{cov}(z(\bfs_1, t_1), z(\bfs_2, t_2) ) 
 = \int_{\mathbb{R}^p} C(\bfh, u; \bfzeta) g^2(\bfeta|\bflambda) \mathrm{d}\bfeta
\end{align}
where $\bfh = \bfs_1 - \bfs_2$ and $u = t_1 - t_2$. Similarly, if the terms involving the kernel function $g$ are normalized with 
$
\tilde{g}(\bfeta| \bflambda) : = {g^2(\bfeta|\bflambda) }/{ \int_{\mathbb{R}^p} g^2(\bfeta|\bflambda) \mathrm{d} \bfeta}, 
$
 the form~\eqref{eqn: cov rep z} is a mixture model representation again.

\subsection{Covariance Function Representation}

To summarize the covariance representation based on the hierarchical model specifications in~\eqref{eqn: process rep} and~\eqref{eqn: process rep z}, we obtain the following hierarchical mixture representation for a class of space-time covariance specifications
\begin{gather} \label{eqn: scale mixture} 
	K(\mathbf{h}, u) \propto \int C(\mathbf{h}, u; \bfeta) \tilde{g}(\bfeta|\bflambda) \mathrm{d} \bfeta, \\ \label{eqn: Lagrangian framework} 
        C(\mathbf{h}, u; \bfzeta) \propto  \int C_s(\mathbf{h}-\mathbf{v}u) \tilde{f}(\bfv|\bfxi) \mathrm{d} \mathbf{v},\\ \label{eqn: velocity}  
    \mathbf{v} | \bfxi \sim \tilde{f}(\mathbf{v} | \bfxi), \quad \bfeta \mid \bflambda \sim \tilde{g}(\bfeta| \bflambda). 
\end{gather}
Here the proportional sign is used because the left-hand side differs from the right-hand side up to some normalizing constants arising from the steps where the kernel functions are normalized into probability density functions. Equation~\eqref{eqn: scale mixture} represents a scale mixture over certain lower-dimensional covariance parameters that could be designed to induce desirable properties such as flexible smoothness behavior and flexible tail decay. Equation~\eqref{eqn: Lagrangian framework} represents a location mixture of the base covariance function $C_s$ over $d$-dimensional probability distribution, which induces space-time asymmetric structure as in the Lagrangian reference framework. Clearly, it also allows spatial anisotropy. The expressions in~\eqref{eqn: velocity} represent the probability distributions for the random vectors $\bfv$ and $\bfeta$. These probability distributions could be designed to vary in space and time so that the space-time process flows over time with the random velocity vector that can vary in space and in time. If desired, a further hierarchical level could be specified for $\bfxi, \bflambda$ to allow more flexible specification. This is similar to a key feature in hierarchical modeling for probability distributions in the sense that additional hierarchical levels can be designed to induce heavy tailed distributions.  

While the hierarchical mixture modeling approach introduced in~\eqref{eqn: scale mixture},~\eqref{eqn: Lagrangian framework} and~\eqref{eqn: velocity} is different from its original usage for probabilistic model specification as covariance functions are not necessarily probability distributions, the concept of hierarchical modeling clearly highlights different levels of mixing operations on covariance functions through the associated parameters and decomposes the complexity of covariance model construction at different levels of hierarchy. The representation of the corresponding space-time processes makes such hierarchical mixture interpretable and easy to construct new models.  

The spatial covariance $C_s$ serves as a base covariance to generating new covariance functions via the Lagrangian reference framework. This work focuses on the base covariance of the form $C_s(\mathbf{h}) = \sigma^2 \exp\{-\|\mathbf{h}\|^2/\rho^2\}, \bfh \in \mathbb{R}^d$. This is often called a squared-exponential (or Gaussian) covariance with the marginal variance $\sigma^2>0$ and the range parameter $\rho>0$. The choice of Gaussian form is motivated by a well-known result \citetext{\citealp{Schoenberg1938}; \citealp[][pp.~44-45]{Stein1999}} that any isotropic correlation functions that are positive-definite in Euclidean space of every dimension can be represented as a scale mixture of Gaussian correlation function over any positive finite measures on $[0, \infty)$. Other choice of base spatial covariance function is also possible, but characterizing the corresponding theoretical properties via Gaussian form brings substantial convenience because of this fundamental result. Throughout this paper, the notation $C_s$ will be reserved to denote this base covariance.  

The notations in Equations~\eqref{eqn: scale mixture},~\eqref{eqn: Lagrangian framework},~\eqref{eqn: velocity} are adopted for uniformity and simplicity of treatment in model construction. The parameterization may not be the most appropriate for inferential or interpretative purposes. When this general construction is discussed within a specific parametric family, these notations will translate into a certain parameterization for inferential or interpretable purposes. 

Given the hierarchical mixture formulation via either covariance representation or space-time process representation, it is desirable to consider a class of models with closed-form expressions from a computational perspective. In terms of practical usefulness, it is also desirable to construct covariance models with flexible behaviors including, but not limited to, degrees of smoothness, tail behaviors (short-range dependence versus long-range dependence), and asymmetry. The resultant covariance models can thus serve as building blocks for modeling real-world applications with possible extensions such as nonstationarity. It seems difficult to establish general conditions on when analytical expressions are available and we shall discuss the practical usefulness of space-time covariance models whenever their analytical expressions are available. This mathematical difficulty could be conveniently leveraged with the hierarchical mixture modeling framework as this approach reformulates a purely integration problem into a hierarchical modeling framework that exploits closed-form derivation and statistical properties at several hierarchical levels.

\section{Main Statistical Properties} \label{sec: property}

We write $f(x) \asymp g(x)$ if $\lim_{x\to \infty} f(x)/g(x) = c\in (0, \infty)$. Further we use the notation $f(x) \sim g(x)$ if $c=1$. We write $X\sim \text{Ga}(a, b)$ if $X$ has a probability density function $p(x) = b^{a}/\Gamma(a) x^{a-1} \exp(-bx),$ $x>0, a>0, b>0$; we write $X\sim \text{IG}(a, b)$ if  $X$ has a probability density function $p(x) = b^{a}/\Gamma(a) x^{-a-1}$ $\exp(-b/x), x>0, a>0, b>0$.

\subsection{Validity of Covariance Models via Hierarchical Mixtures}

Given the covariance function specifications in Equations~\eqref{eqn: cov rep} and~\eqref{eqn: cov rep z}, a natural question is whether they are positive definite (or at least positive semi-definite) when their mixing measures vary in space and time. This is addressed in the following result with its proof given in Appendix~\ref{app: pd}. 

\begin{theorem} \label{thm: pd}
Suppose that $\mu$ is a nonnegative measure on $\mathbb{R}^d$, $\nu$ is a nonnegative measure on $\mathbb{R}^p$, and $C_s$ is a stationary covariance function in $\mathbb{R}^d$. 
For any fixed $\Ell:=(\mathbf{s}, t) \in \mathbb{R}^d\times \mathbb{R}$, let $f(\cdot | \bfxi_{\Ell}) \in L^2(\mu): \mathbb{R}^d \to \mathbb{R}$ denote a kernel function whose hyperparameters may be a function of $\Ell$. Let $g(\cdot | \bfeta_{\Ell}): \mathbb{R}^p \to \mathbb{R}$ denote an integrable  kernel whose hyperparameters may be a function of $\Ell$. For any space-time locations $\Ell_1, \Ell_2 \in \mathbb{R}^d\times \mathbb{R}$,  define the function 
\begin{align*}
C(\Ell_1, \Ell_2; \bfzeta) := \int_{\mathbb{R}^d} C_s(\bfh - \bfv u) f(\bfv| \bfxi_{\Ell_1}) f(\bfv|\bfxi_{\Ell_2}) \mu(\mathrm{d} \bfv)
\end{align*}
where $\bfh:=\bfs_1 - \bfs_2$, $u:=t_1 - t_2$, $\bfzeta$ denotes the vector of parameters in both $C_s$ and $f$. Similarly, for a subvector $\bfeta\in \mathbb{R}^p$ of $\bfzeta$, define the function 
\begin{align*}
    K(\Ell_1, \Ell_2):= \int_{\mathbb{R}^p} C(\Ell_1, \Ell_2; \bfzeta) g(\bfeta|\bflambda_{\Ell_1}) g(\bfeta|\bflambda_{\Ell_2}) \nu(\mathrm{d}\bfeta).
\end{align*} 
Then both $C$ and $K$ are valid covariance functions provided that $K$ exists. 
\end{theorem}
\begin{remark}
This result indicates that the hierarchical mixture based functions $C$ and $K$ above are valid covariance functions even if their mixing distributions vary in space-time domain. Thus, the above covariance functions are generalizations of the stationary covariance functions in~\eqref{eqn: cov rep} and~\eqref{eqn: cov rep z}. Nonstationarity could be achieved when either $f$ or $g$ varies in space-time domain. The positive definiteness in~\eqref{eqn: cov rep} and~\eqref{eqn: cov rep z} is guaranteed. 
\end{remark}

\subsection{Smoothness and Tail Behaviors} 
A key benefit of using a Gaussian covariance as the base covariance is that  the behaviors of the resulting covariance functions near the origin and the tails can be characterized via the following result with its proof given in Appendix~\ref{app: tail behavior}. 
\begin{theorem} \label{thm: tail behavior} 
Let $U$ be a strictly positive random variable with probability measure $G$. 
Then the scale mixture of Gaussian correlation functions 
 \begin{align*} %\label{eqn: univariate SM} 
 R(\bfh) :=  \int_0^{\infty} \exp\left\{- \frac{\|\bfh\|^2}{2u} \right\} \mathrm{d} G(u), \quad \mathbf{h} \in \mathbb{R}^d,
 \end{align*}
 is an isotropic covariance function on $\mathbb{R}^d$ with the following properties. 
\begin{itemize}[nosep]
\item[(i)]
A Gaussian spatial process with covariance function $R(\cdot)$ is $m$-times mean-square differentiable if the $m$-th moment of the random variable $1/U$ exists. 
\item[(ii)] Assume that $U$ has a continuous density $g$ on $(0, \infty)$. 
Let $L(x)$ denote a slowly varying function at $\infty$ such that $\lim_{x\to \infty} L(tx) / L(x) = 1$ for all $t \in (0, \infty)$.
Assume that the tail of the mixing density $g$ satisfies 
\begin{align*}
g(u) \asymp \exp(-c_0u) u^{\lambda-3/2} L(u), \text{ as } u\to \infty,
\end{align*}
where $c_0$ is a nonnegative constant. 
If $c_0=0$, then
$
R(\bfh) \asymp \|\bfh\|^{2\lambda -2} L(\|\bfh\|^2)  \text{ as } \|\bfh\| \to \infty;  
$
If $c_0>0$, then 
$
R(\bfh) \asymp \|\bfh\|^{\lambda -3/2} L(\|\bfh\|) \exp\{- (2c_0)^{1/2} \|\bfh\| \} \text{ as } \|\bfh\| \to \infty.
$

\end{itemize}

\end{theorem}  
\begin{remark}
The above theorem says that (i) the smoothness of covariance is determined by the existence of the moment of the (inverse of) scale mixing random variable; (ii) if the mixing density has a polynomial tail, so does the resulting covariance function; if the mixing density has an exponential tail, so does the resulting covariance function. Power-law covariances with long-range dependence can be constructed via mixing densities with polynomial tails. This property can be verified with the following spatial covariance models.
\end{remark}

% \begin{example}[The power-exponential class]
%  Let $1/U$ denote the positive stable random variable with index $\alpha/2 \in (0, 1)$ with its density denoted as $f_{\alpha/2}$. Then $K$ is the power-exponential correlation function of the form \citep[e.g.,][]{Gneiting1997} 
% \begin{align*}
%     K(\mathbf{h})  =  \exp\left\{ - \left(\frac{\|\mathbf{h}\|}{\phi}\right)^{\alpha}  \right\}, 
% \end{align*}
% where $\alpha \in (0, 2)$ and $\phi>0$. 
% \begin{proof}
% As $1/U \sim f_{\alpha/2}$, we have $E[\exp(-r/U)]=\exp(-r^{\alpha/2})$ for $r\geq 0$. Let $V=1 / (\phi^2 U) $. Then 
% \begin{align*}
%     \exp\left\{ - \left(\frac{\|\mathbf{h}\|}{\phi}\right)^{\alpha}  \right\} = E[\exp(-V \|\bfh\|^2)] = \int_0^{\infty} \exp(-v\|\bfh\|^2)  f_V(v) \text{d} v. 
% \end{align*}
% where $f_V$ denotes the pdf of $V$.
% \end{proof}
% \end{example}

% \begin{example}[The Cauchy class]
% Let $U\sim \mathcal{IG}(\nu, \phi^2/2)$ with density 
% $$p(u) = \frac{\phi^{2\nu}}{ 2^{\nu} \Gamma(\nu)} u^{-\nu-1} \exp\left(- \frac{\phi^2}{2u}\right).$$ 
% Then $K$ is the Cauchy correlation function of the form 
% \begin{align*}
% K(\bfh) = \left( 1 + \frac{\|\bfh\|^2}{\phi^2} \right)^{-\nu},
% \end{align*}
% where  $\phi>0$ is the range parameter, and $\nu$ is the smoothness parameter.    
% \end{example}

\begin{example}[The Mat\'ern class] \label{ex: Matern}
Let $U\sim \text{Ga}(\nu, 1/(2\phi^2))$.  
Then  $K$ is the Mat\'ern correlation function of the form 
\begin{align*} 
\calM(\bfh; \nu, \phi) = \frac{2^{1-\nu}}{\Gamma(\nu)} \left( \frac{\|\bfh\|}{\phi} \right)^{\nu} \calK_{\nu}\left( \frac{\|\bfh\|}{\phi} \right),\quad \bfh \in \mathbb{R}^d,
\end{align*}
where  $\phi>0$ is the range parameter, $\nu$ is the smoothness parameter, and $\calK_{\nu}$ is the modified Bessel function of the second kind of order $\nu$. As $\nu\to \infty$, the Mat\'ern covariance becomes a Gaussian covariance. Applying Theorem~\ref{thm: tail behavior} yields $M(\bfh; \nu, \phi) \asymp \|\bfh/\phi \|^{\nu-1/2} \exp(-\|\bfh\|/\phi)$ as $\|\bfh \|\to \infty$. 
\end{example} 

\begin{example}[The CH class]
%Of particular interest is the construction of the Confluent Hypergeometric (CH) class. Indeed, 
Let $U\sim\frac{\beta^2}{2} \text{IB}(\nu,\alpha)$ follow a scaled inverted beta (or beta prime) distribution with density 
\begin{align*}
p(u) =  \frac{2}{\beta^2 B(\nu, \alpha)} \left( \frac{2u}{\beta^2} \right)^{\nu-1}  \left( 1 + \frac{2u}{\beta^2} \right)^{-\nu-\alpha},\quad u>0. 
\end{align*}
Then $K$ is the Confluent Hypergeometric (CH) correlation function of the form
\begin{align*}
 \mathcal{CH}(\bfh; \nu, \alpha, \beta) =  \frac{ \Gamma(\nu+\alpha)}{\Gamma(\nu)} \calU\left(\alpha, 1-\nu,  \left(\frac{\|\bfh\|}{\beta}\right)^2\right), \quad \mathbf{h} \in \mathbb{R}^d,
\end{align*}
where $\alpha>0$ is the tail decay parameter, $\beta>0$ is the range parameter, $\nu>0$ is the smoothness parameter, and $\calU$ is the confluent hypergeometric function of the second kind. As $\nu\to \infty$, the CH correlation becomes the Cauchy correlation \citep{Ma2023}. It follows from Theorem~\ref{thm: tail behavior}  that $\mathcal{CH}(\bfh; \nu, \alpha, \beta) \sim  \beta^{2\alpha} \Gamma(\nu+\alpha) / \Gamma(\nu) \|\bfh\|^{-2\alpha} L(\|\bfh\|^2)$ as $\bfh\to \infty$, where $L(x)= \{ x / (x + \beta^2/2)\}^{\nu+\alpha}$. 
\end{example}

\section{A New Class of Asymmetric Space-Time Covariance Models} \label{sec: stcov}

\subsection{A Lagrangian Mat\'ern Model} \label{sec: LagMatern}
\begin{figure}[!ht]
\begin{center}
\makebox[\textwidth][c]{\includegraphics[width=1\textwidth, height=0.45\textheight]{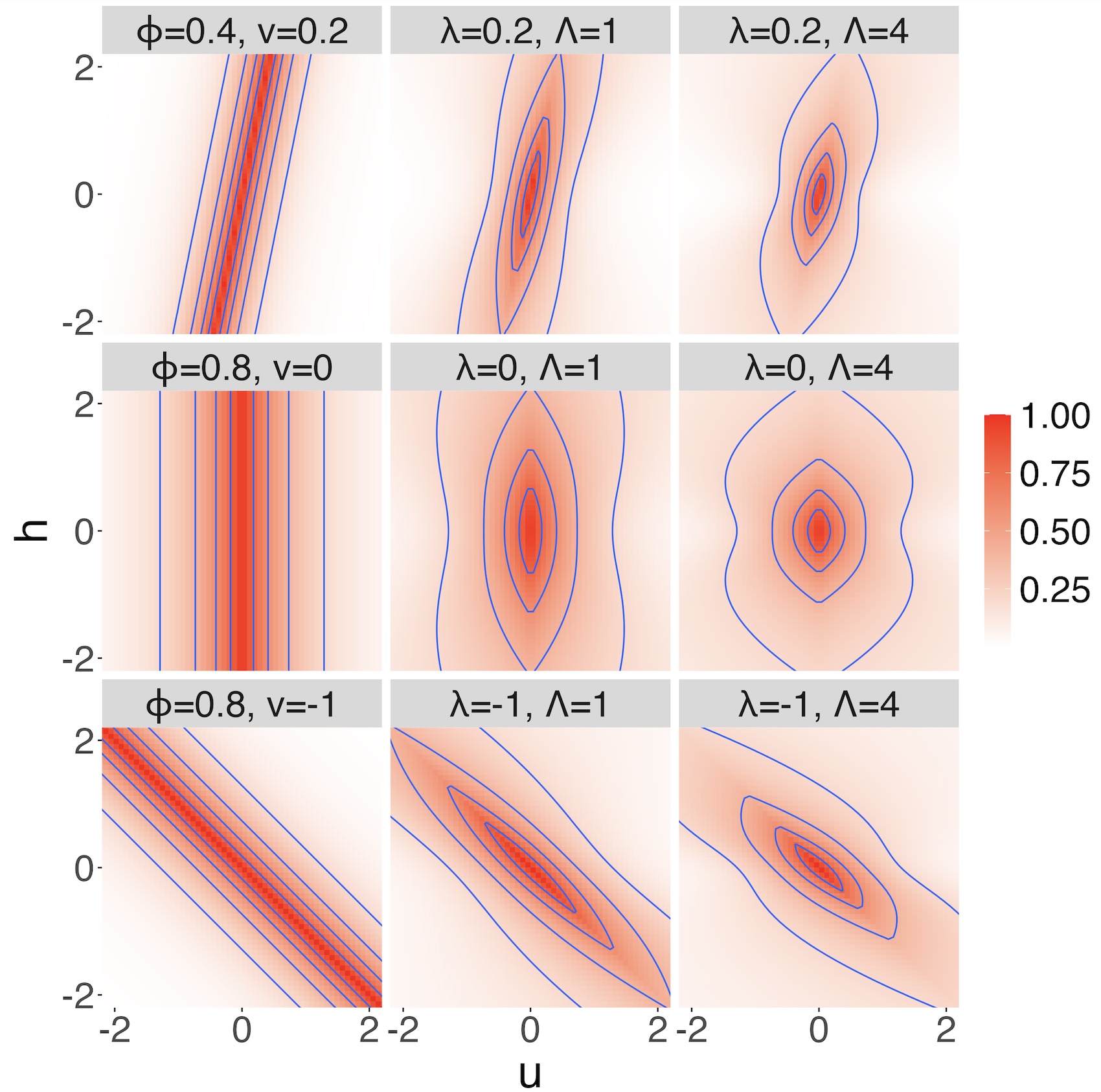}}
\caption{Contour plots of asymmetric space-time Mat\'ern models over the space-time domain $[-2,2]\times [-2, 2]$. Horizontal axis represents spatial lag and vertical axis represents temporal lag. The first column corresponds to the frozen asymmetric Mat\'ern models with different parameter settings. The second and third columns correspond to the (non-frozen) Lagrangian Mat\'ern models with different parameter settings for $\bflambda$ and $\bfLambda$. All panels share the same Mat\'ern parameters $\sigma^2=1, \nu=0.5$. The range parameter $\phi$ is the same across each row. The other parameters are specified in the title of each panel. \vspace{-.6cm}}
\label{fig: LMatern contours}
\end{center}
\end{figure}

The first class of new nonseparable space-time covariance models is to consider the following hierarchical model specification: 
\begin{gather*} %\label{eqn: LagMatern model}
    %\begin{split}
        K(\mathbf{h}, u) = \int C(\mathbf{h}, u; \rho^2) \text{d} G(\rho^2), \\
        C(\mathbf{h}, u; \rho^2) = E_{\mathbf{v}}[C_s(\mathbf{h}-\mathbf{v}u)], \\
        \mathbf{v} | \rho, \bflambda, \bfLambda \sim \mathcal{N}_d(\bflambda, \rho^2 \bfLambda/2), \quad
        \rho^2 | \phi^2 \sim \text{Ga}(\nu, 1/(4\phi^2)), 
    %\end{split}
\end{gather*}
where $\nu>0, \phi>0$, $\bflambda \in \mathbb{R}^d$, and $\bfLambda$ is a $d\times d$ positive-definite matrix. $G$ denotes the probability measure for $\rho^2$ which is specified by the Gamma density. A key advantage of this model specification is that an analytical form is available as shown in Proposition~\ref{prop: LagMatern} with its proof given in Appendix~\ref{app: LMatern}. Besides, after mixing with $\bfv$, we obtain a Gaussian form which allows the application of Theorem~\ref{thm: tail behavior}. The contour plots of the covariance function under different parameter settings are given in Figure~\ref{fig: LMatern contours}. 

\begin{proposition} \label{prop: LagMatern}
The hierarchical specification above yields 
\begin{align} \label{eqn: LagMatern cov}
   K(\mathbf{h}, u) =  \sigma^2 |\mathbf{I}_{d\times d} + u^2 \bfLambda|^{-1/2} \mathcal{M}(r_u; \nu, \phi)
\end{align}
where $r_u:=\{(\mathbf{h} - u\bflambda)^\top (u^2 \bfLambda + \mathbf{I}_{d\times d})^{-1} (\mathbf{h} - u\bflambda)\}^{1/2}$. $\sigma^2>0$ is the variance parameter, $\phi>0$ is the range parameter, $\nu>0$ is a smoothness parameter, $\bflambda \in \mathbb{R}^d$ is the (prior) velocity vector and $\bfLambda \in \mathbb{R}^{d\times d}$ is the (prior) covariance matrix of the velocity vector that allows geometric anisotropy. We call this model a parsimonious \textit{Lagrangian Mat\'ern}  model due to its inclusion of the Mat\'ern correlation function. 
\end{proposition}

\begin{remark}
A nice property of this covariance class is that it has flexible origin behavior controlled by the smoothness parameter $\nu$ that allows arbitrary different degrees of smoothness. Since the term $|\mathbf{I}_{d\times d} + u^2 \bfLambda|^{-1/2}$ is analytic in terms of $u$, the smoothness in time and in space is solely controlled by $\nu$ when $\bflambda \neq \bf0$. Another close construction leads to the so-called Lagrangian hyperbolic model given in Appendix~\ref{sec: LagGH}  with the same flexibility in terms of smoothness and tail behaviors as the  Lagrangian Mat\'ern model. 
\end{remark}

\begin{remark}
As $\|\bfh\| \to \infty$, for any fixed $u$, $K(\bfh, u)$ decays exponentially fast as in the Mat\'ern class. However, as $u\to \infty$, for any fixed $\bfh$, $K(\mathbf{h}, u) \asymp u^{-d}$, which indicates that the Lagrangian Mat\'ern covariance has algebraic decay in time. If $d=1$, $K(\bfh, u)$ is not integrable in $u$ and hence allows long-range dependence in time; for integer $d>1$, $K(\bfh, u)$ is integrable in time and hence does not allow long-range dependence in time.
\end{remark}

\begin{remark}
If the spatial domain is in $\mathbb{R}^2$, we can parameterize the prior velocity vector $\bflambda$ and covariance 
$\bfLambda$: $\bflambda = \lambda_0 [\cos(\theta_0), \sin(\theta_0)]^\top$, $\bfLambda:= \mathbf{U} \diag\{\lambda_1, \lambda_2 \} \mathbf{U}^\top$, where $\lambda_0>0, \theta_0\in(-\pi, \pi]$, $\lambda_1>0, \lambda_2>0$ and 
$$
\mathbf{U} = \begin{bmatrix}
    \cos(\theta_1) & - \sin(\theta_1) \\
    \sin(\theta_1) & \cos(\theta_1)
\end{bmatrix}
$$ 
is a 2-by-2 rotation matrix with $\theta_1\in (-\pi, \pi]$. This parameterization is adopted in numerical illustrations for likelihood-based inference. 
\end{remark}

\begin{remark}
This new class of covariance models includes the Lagrangian Gaussian model~\eqref{eqn: LagGauss Cov} as a special case when $\nu\to \infty$. As in the Mat\'ern class, when $\nu$ takes half integers, analytical forms are available as well. We give an example below without proofs as their derivations follow trivially by applying the properties of the modified Bessel function.  
\end{remark}

 \begin{example}[Space-Time Mat\'ern with $\nu=1/2$]
When $\nu=1/2$, the Mat\'ern correlation function becomes an exponential correlation function. Thus, 
$
    K(\bfh, u) = \sigma^2 | \bfI_{d\times d} + u^2 \bfLambda|^{-1/2} \exp ( -  {r_u}/{\phi})
$
where $r_u:=\{(\mathbf{h} - u\bflambda)^\top (u^2 \bfLambda + \mathbf{I}_{d\times d})^{-1} (\mathbf{h} - u\bflambda)\}^{1/2}$.
 \end{example}

\begin{remark}
When $\bflambda = \mathbf{0}$, the covariance function~\eqref{eqn: LagMatern cov} is nonseparable with fully symmetric structures. When $\bflambda=\mathbf{0}, \bfLambda=a\mathbf{I}_{d\times d}$ for any $a>0$, it can be shown that the covariance function~\eqref{eqn: LagMatern cov} is an example of Gneiting's covariance class \citep[][Eq.~(16)]{Gneiting2002}.
\end{remark}

%\begin{remark}
%If there is a need to allow more flexible control over smoothness  in space and in time, one can easily achieve this by taking $\rho^2 \sim \text{Ga}(\nu(\bfh, u), 1/(4\phi^2))$, where $\nu(\bfh, u)$ varies in space and time. The resulting function is guaranteed to be a positive-definite covariance function by Theorem~\ref{thm: pd}. 
%\end{remark}

\subsection{A Lagrangian CH Model} \label{sec: LagCH}
\begin{figure}[!ht]
\begin{center}
\makebox[\textwidth][c]{\includegraphics[width=1\textwidth, height=0.45\textheight]{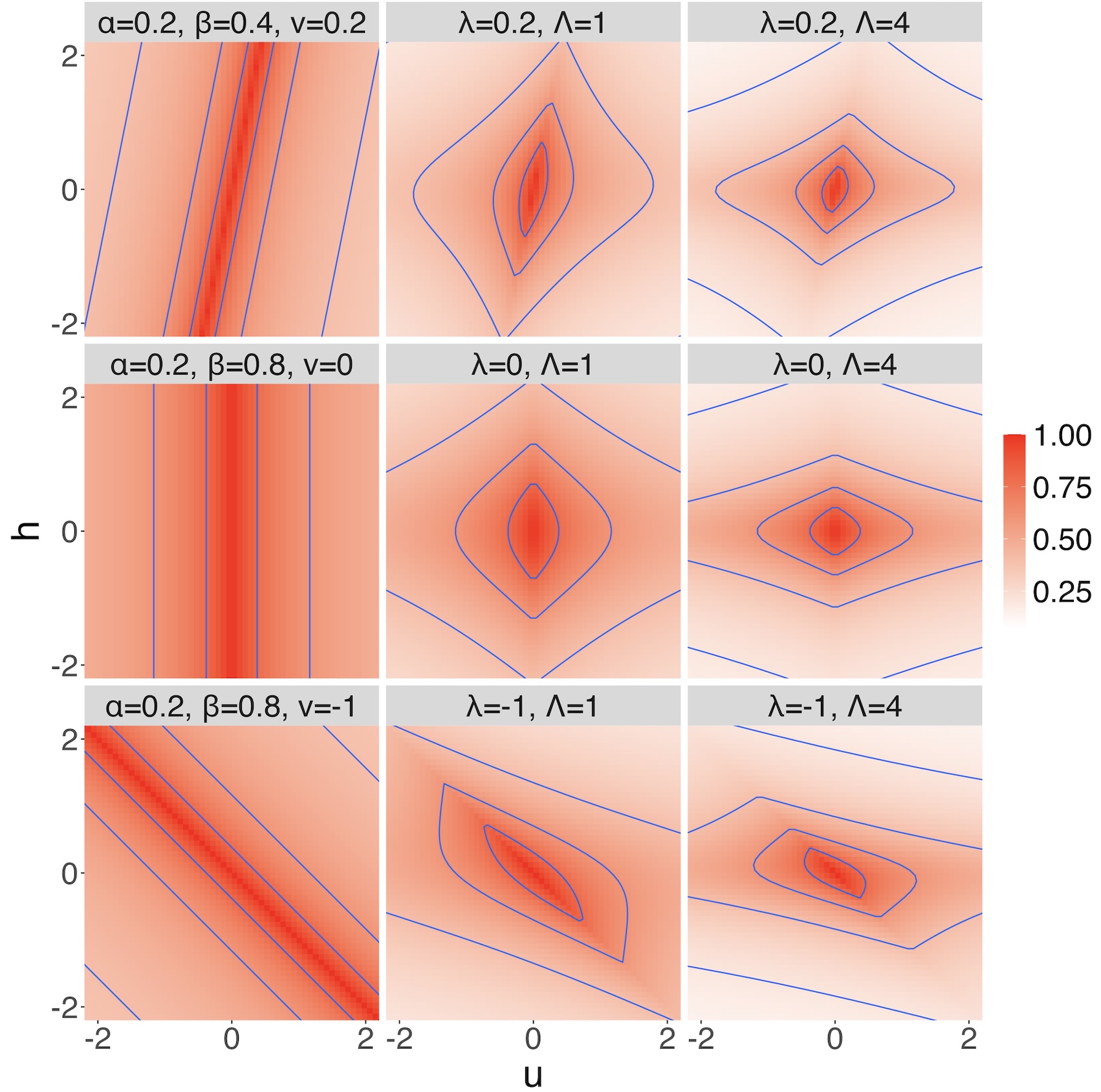}}
\caption{Contour plots of asymmetric CH models over the space-time domain $[-2,2]\times [-2, 2]$. The first column corresponds to the frozen asymmetric CH models. The second and third columns correspond to the (non-frozen) Lagrangian CH models with different parameter settings for $\bflambda$ and $\bfLambda$. All panels share the same parameters $\sigma^2=1, \nu=0.5, \alpha=0.2$. Each row shares the same range parameter $\beta$ and tail decay parameter $\alpha$. All other parameter settings are specified in the title of each panel. \vspace{-.6cm}}
\label{fig: LCH contours}
\end{center}
\end{figure}

One property of the Mat\'ern covariance function is that it has exponential decay in the tail, which is also preserved in the Lagrangian Mat\'ern class developed in Section~\ref{sec: LagMatern}. Exploiting Theorem~\ref{thm: tail behavior}, we can generate covariance functions with polynomial decay to allow long-range dependence. To see this, we can introduce an additional hierarchy of mixture over $\phi^2$ with the following hierarchical model specification: 
\begin{gather*} %\label{eqn: LagCH model}
    %\begin{split}
        K(\mathbf{h}, u) = \int C(\mathbf{h}, u; \rho^2, \phi^2) \text{d} G(\rho^2, \phi^2),\quad   \\
        C(\mathbf{h}, u; \rho^2, \phi^2) = E_{\mathbf{v}}[C_s(\mathbf{h}-\mathbf{v}u)], \\
        \mathbf{v} | \rho^2, \bflambda, \bfLambda \sim \mathcal{N}_d(\bflambda, \rho^2 \bfLambda/2), \,
        \rho^2 | \phi^2 \sim \text{Ga}(\nu, 1/(4\phi^2)), \,
        \phi^2 | \alpha, \beta^2 \sim \text{IG}(\alpha, \beta^2/4),
    %\end{split}
\end{gather*}
where $\alpha>0, \beta>0, \nu>0$, $\bflambda \in \mathbb{R}^d$, and $\bfLambda$ is a $d\times d$ positive-definite matrix.  $G$ is the joint probability measure specified by $p(\rho^2|\phi) p(\phi^2)$. 
As expected, a closed-form expression is given in Proposition~\ref{prop: LagCH} with its proof given in Appendix~\ref{app: LCH}. The contour plots of the covariance function under different parameter settings are given in Figure~\ref{fig: LCH contours}. 
\begin{proposition} \label{prop: LagCH}
The above model specification yields 
\begin{align} \label{eqn: LagCH cov}
   K(\mathbf{h}, u) =  \sigma^2 |I_{d\times d} + u^2 \bfLambda|^{-1/2} \mathcal{CH}(r_u; \nu, \alpha, \beta)
\end{align}
where $r_u:=\{(\mathbf{h} - u\bflambda)^\top (u^2 \bfLambda + \mathbf{I}_{d\times d})^{-1} (\mathbf{h} - u\bflambda)\}^{1/2}$.
$\sigma^2>0$ is the variance parameter, $\alpha>0$ is the tail decay parameter, $\beta>0$ is the range parameter, $\nu>0$ is a smoothness parameter, $\bflambda \in \mathbb{R}^d$ is the (prior) velocity vector and $\bfLambda \in \mathbb{R}^{d\times d}$ is the (prior) covariance matrix of the velocity vector. 
We call this model a parsimonious \textit{Lagrangian CH} model due to its inclusion of the CH correlation function.
\end{proposition}

\begin{remark}
This covariance function builds on top of the Lagrangian Mat\'ern model by adding an additional level of scale mixing, which preserves origin behavior but inflates the tail. As $\|\bfh\|\to \infty$, for any fixed $u$, $K(\bfh, u) \asymp \sigma^2 |I_{d\times d} + u^2 \bfLambda|^{-1/2} \beta^{2\alpha} / \Gamma(\nu) r_u^{-2\alpha} L(r_u^2)$, where $L(x) = \{x / (x + \beta^2/2)\}^{\nu+\alpha}$. This indicates that it decays algebraically in space. $K(\bfh, u)$ is not integrable in $\bfh$ for $\alpha \in (0, d/2]$ and hence  allows long-range dependence. As $u\to \infty$, for any fixed $\bfh$, $K(\bfh, u) \asymp \sigma^2 |\bfLambda|^{-1/2} \beta^{2\alpha} u^{-d} r_u^{-2\alpha} L(r_u^2)$ for $\bflambda\neq 0$, which indicates that  $K(\bfh, u)$ also has algebraic decay in time.  The Lagrangian CH covariance function thus can allow long-range dependence in time for $d=1$.
\end{remark}

\subsection{Extension to Nonstationary Lagrangian Models} \label{label: Lag nonstationarity}
This hierarchical mixture approach can be used to introduce Lagrangian based space-time nonstationarity where the nonstationarity arises due to the space-time varying velocity field. Let $\bfv(\cdot):=\{\bfv(\Ell):=(v_1(\Ell), \ldots, v_d(\Ell))^\top \in \mathbb{R}^d: \Ell  \in \mathbb{R}^{d+1} \}$ denote a $d$-variate space-time process with vector-valued mean function $\bflambda(\cdot): \mathbb{R}^{d+1}\mapsto \mathbb{R}^d$ and matrix-valued covariance function $\rho^2\bfLambda(\cdot, \cdot)/2: \mathbb{R}^{d+1}\times \mathbb{R}^{d+1} \mapsto \mathbb{R}^{d\times d}$. Given $\rho^2$ and $\bfv$,  the space-time process $\{w(\bfs - t \bfv(\bfs, t)): (\bfs, t)\in \mathbb{R}^{d+1}\}$ has the covariance function $C(\Ell_i, \Ell_j; \rho^2, \bfv):=\text{cov}\{w(\bfs_i - t_i \bfv(\Ell_i)), w(\bfs_j - t_j\bfv(\Ell_j) )| \rho^2, \bfv\}$.  The following result shows that the hierarchical mixture constructions in Sections~\ref{sec: LagMatern} and~\ref{sec: LagCH} yield the nonstationary Lagrangian Mat\'ern and CH models. 

\begin{proposition} \label{prop: Lag nonstationarity} 
Suppose that $\bfv(\cdot) | \rho^2 \sim \mathcal{MGP}(\bflambda(\cdot), \rho^2\bfLambda(\cdot,\cdot)/2)$ is a $d$-variate Gaussian space-time process.  For any $\Ell_i, \Ell_j\in \mathbb{R}^{d+1}$, consider the following hierarchical mixture
\begin{gather*}
K(\Ell_i, \Ell_j): =  E_{\phi^2}[C(\Ell_i, \Ell_j; \phi^2)],\quad C(\Ell_i, \Ell_j; \phi^2) := E_{\rho^2} [C(\Ell_i, \Ell_j; \rho^2) | \phi^2], \\ 
C(\Ell_i, \Ell_j; \rho^2):= E_{\bfv}[C(\Ell_i, \Ell_j; \rho^2, \bfv)] \\ 
\rho^2 | \phi^2 \sim \mathrm{Ga}(\nu, 1/(4\phi^2)), \, \phi^2|\alpha, \beta^2 \sim \mathrm{IG}(\alpha, \beta^2/4).
\end{gather*} 
Then $C(\Ell_i, \Ell_j; \phi^2) =  \sigma^2 |\bfO_{ij}|^{-1/2} \mathcal{M}(r_{ij}; \nu, \phi)$ and $K(\Ell_i, \Ell_j)=\sigma^2 |\bfO_{ij}|^{-1/2} \mathcal{CH}(r_{ij}; \nu, \alpha, \beta)$, 
where $\bfQ_{ij} :=t_i^2 \bfLambda(\Ell_i, \Ell_i) + t_j^2 \bfLambda(\Ell_j, \Ell_j) - t_i t_j [\bfLambda(\Ell_{i}, \Ell_j) + \bfLambda(\Ell_j, \Ell_i)]$ and  $r_{ij}^2:=(\bfh_{ij} - \bfm_{ij})^\top \bfO_{ij}^{-1} (\bfh_{ij} - \bfm_{ij})$   with $\bfh_{ij}= \bfs_i - \bfs_j$,  $\bfO_{ij}:=\bfI_{d\times d} + \bfQ_{ij}$ and $\bfm_{ij} := t_i \bflambda(\Ell_i) - t_j\bflambda(\Ell_j)$.
\end{proposition} 

\begin{remark}
The hierarchical mixture construction  can allow various types of nonstationarity in space and in time even if a stationary velocity field is assumed; see Appendix~\ref{app: nonstationarity} for various illustrations. The resulting nonstationarity driven by a velocity field that varies in space and time is physically motivated. Moreover, this hierarchical mixture approach provides a way to integrate flow information through velocity modeling in covariance construction. 
\end{remark}

\begin{remark}
If it is desirable to have more flexible behavior over smoothness  in space and in time, one can achieve this by taking $\rho^2 \sim \text{Ga}(\nu(\bfh, u), 1/(4\phi^2))$, where $\nu(\bfh, u)$ varies in space and time. Similarly, to allow more flexible control over algebraic decay in space and in time, one could take $\phi^2 \sim \text{IG}(\alpha(\bfh, u), \beta^2/4)$ with a spatio-temporally varying function $\alpha(\bfh, u)$; see \citet{Chen2026} for various examples. 
\end{remark}

\subsection{Extension to Multivariate Lagrangian Models}
To extend to the multivariate case, it is crucial to choose appropriate mixing measures for different space-time processes. 
Consider a $q$-variate spatial process with base cross-covariance $C_{ij}^0(\bfh): = \sigma_i \sigma_j a_{ij} \exp(-\|\bfh\|^2/\rho^2)$, where $\sigma_i>0$ and $\bfA=[a_{ij}]$ is assumed to be a $q\times q$ positive-definite correlation matrix with $a_{ii}=1$, and hence the matrix-valued covariance function $[C_{ij}^0(\cdot,\cdot)]$ is positive definite. 
Let $\bfV:=[\bfv_1^\top(\cdot), \ldots, \bfv_q^\top(\cdot)]^\top \sim N_{qd}(\bfmu_{\bflambda}, \rho^2\bfSigma_{\bfLambda} /2)$, where  $\bflambda_i:=E[\bfv_i] \in \mathbb{R}^d$ and $\text{cov}(\bfv_i, \bfv_j)=\bfLambda_{ij} \rho^2/2$. Define $C_{ij}(\Ell_k, \Ell_m; \bfV, \rho^2) := C_{ij}^0(\bfs_k - \bfs_m - t_k \bfv_i + t_m \bfv_j)$. The following multivariate hierarchical mixture construction provides valid multivariate Lagrangian models. 

%Let $\bfY(\cdot) :=[y_1(\cdot), \ldots, y_q(\cdot)]^\top$
%be a $q$-variate space-time process transported from $\bfw(\cdot)$ componentwise by $\bfV$. That is, given $\bfV$,  
%$y_i(\bfs, t) := w_i(\bfs - \bfv_i t)$ for $i=1,\ldots, q$. Then for $\Ell_k=(\bfs_k, t_k),\Ell_m=(\bfs_m, t_m)$, the conditional cross-covariance of $\bfY(\cdot)$ is $C_{ij}(\Ell_k, \Ell_m; \bfV, \rho^2):= \mathrm{cov}\{y_i(\Ell_i), y_j(\Ell_j) | \bfV, \rho^2\}$. 

%Let $p_i(\rho^2)=\mathrm{Ga}(\rho^2| \nu_i, 1/(4\phi^2))$ and $p_0(\rho^2):= (1/q) \sum_{k=1}^q p_i(\rho^2)$.  
%Define the scale-mixing process $z_i(\cdot) := \sqrt{p_i(\rho^2) / p_0(\rho^2)} y_i(\cdot)$ for $i=1,\ldots, q$ with cross-covariance $K_{ij}(\cdot, \cdot; \rho^2)$. Similarly, let $\pi_i(\phi^2) = \text{IG}(\phi^2 | \alpha_i, \beta_i^2/4)$ and $\pi_0(\phi^2) = (1/q) \sum_{k=1}^q \pi_i(\phi^2)$. Define the scale-mixing process $\tilde{z}_i(\cdot):=\sqrt{\pi_i(\phi^2) / \pi_0(\phi^2)} z_i(\cdot)$ for $i=1,\ldots, q$, with cross-covariance $\tilde{K}_{ij}(\cdot, \cdot)$.  The following result derives their analytic forms through a hierarchical mixture construction.  

\begin{proposition}[Multivariate hierarchical mixtures]
\label{prop: mv mixture}
For  any positive integers $q,d$,   let $C_{ij}(\Ell_k,$ $\Ell_m;\rho^2) :=E_{\bfV\mid\rho^2} \{C_{ij}(\Ell_k,\Ell_m;\bfV,\rho^2)\}$. 
Define
$
\bfh_{ij}=\bfs_k-\bfs_m-t_k\bflambda_i+t_m\bflambda_j,\, 
\bfQ_{ij} =t_k^2\bfLambda_{ii}+t_m^2\bfLambda_{jj}  -t_kt_m(\bfLambda_{ij}+\bfLambda_{ji}),\,
\bfO_{ij} =\mathbf I_{d\times d}+\bfQ_{ij}, \, r_{ij}^2=\bfh_{ij}^{\top}\bfO_{ij}^{-1}\bfh_{ij},
$
and define
\[
\nu_{ij}=\frac{\nu_i+\nu_j}{2},
\qquad
\phi_{ij}^2=\frac{2}{\phi_i^{-2}+\phi_j^{-2}},
\qquad
\alpha_{ij}=\frac{\alpha_i+\alpha_j}{2},
\qquad
\beta_{ij}^2=\frac{\beta_i^2+\beta_j^2}{2}.
\]

\noindent\textnormal{(i) Multivariate Mat\'ern Model.}
For any $\nu_i,\phi_i>0$, let
$
p_i(\rho^2)
=\mathrm{Ga}\left(\rho^2| \nu_i, {1}/{(4\phi_i^2)}\right),\,
p_0(\rho^2):=\frac1q\sum_{i=1}^q p_i(\rho^2).
$
Under $\rho^2\sim p_0$,  
\begin{align*}
K_{ij}(\Ell_k,\Ell_m)
&:= E_{p_0}\left[\frac{\sqrt{p_i(\rho^2)p_j(\rho^2)}}{p_0(\rho^2)} C_{ij}(\Ell_k,\Ell_m;\rho^2)\right] 
= \sigma_i\sigma_j a_{ij}c_{ij}^{\nu} \frac{\phi_{ij}^{2\nu_{ij}}}{\phi_i^{\nu_i}\phi_j^{\nu_j}} |\bfO_{ij}|^{-1/2}\mathcal M(r_{ij};\nu_{ij},\phi_{ij}).
\end{align*}
and $\bfK(\cdot, \cdot)=[K_{ij}(\cdot, \cdot)]$ is a valid multivariate Lagrangian Mat\'ern model.

\noindent\textnormal{(ii) Multivariate CH model.}
For any $\nu_i,\alpha_i,\beta_i>0$, define 
\begin{gather*}
p_i(\rho^2\mid\phi^2)=\mathrm{Ga}\left(\rho^2| \nu_i, {1}/{(4\phi^2)}\right),\, \quad
p_0(\rho^2\mid\phi^2)=\frac1q\sum_{i=1}^q p_i(\rho^2\mid\phi^2),\\  
\pi_i(\phi^2)=\mathrm{IG}\left(\phi^2| \alpha_i, {\beta_i^2}/{4}\right),\, \quad
\pi_0(\phi^2)=\frac1q\sum_{i=1}^q\pi_i(\phi^2).
\end{gather*}
Under the hierarchy
\[
\phi^2\sim\pi_0,
\qquad
\rho^2\mid\phi^2\sim p_0(\cdot\mid\phi^2),
\]
and  
\begin{gather*}
K_{ij}(\Ell_k,\Ell_m;\phi^2)
:=
E_{p_0(\cdot\mid\phi^2)}\left[
\frac{\sqrt{p_i(\rho^2\mid\phi^2)p_j(\rho^2\mid\phi^2)}}
     {p_0(\rho^2\mid\phi^2)}
C_{ij}(\Ell_k,\Ell_m;\rho^2)
\right], 
\end{gather*}
it follows that 
\begin{gather*}
\tilde K_{ij}(\Ell_k,\Ell_m)
:=
E_{\pi_0}\left[
\frac{\sqrt{\pi_i(\phi^2)\pi_j(\phi^2)}}{\pi_0(\phi^2)}
K_{ij}(\Ell_k,\Ell_m;\phi^2)
\right]
=\sigma_i\sigma_j a_{ij}c_{ij}^{\nu}c_{ij}^{\alpha,\beta}
|\bfO_{ij}|^{-\frac{1}{2}}
\mathcal{CH}(r_{ij};\nu_{ij},\alpha_{ij},\beta_{ij}).
\end{gather*}
and $\tilde{\bfK}(\cdot, \cdot)=[\tilde K_{ij}(\cdot,\cdot)]$ is a valid multivariate Lagrangian CH covariance. In both cases,
\begin{gather*}
c_{ij}^{\nu}
=
\frac{\Gamma(\nu_{ij})}
     {\sqrt{\Gamma(\nu_i)\Gamma(\nu_j)}},
\qquad
c_{ij}^{\alpha,\beta}
=
\frac{\Gamma(\alpha_{ij})}
     {\sqrt{\Gamma(\alpha_i)\Gamma(\alpha_j)}}
\frac{\beta_i^{\alpha_i}\beta_j^{\alpha_j}}
     {\beta_{ij}^{2\alpha_{ij}}}.
\end{gather*}
\end{proposition}
\begin{remark}
In the above parsimonious construction, the space-time multivariate models are valid for any integer $q>1$, and all the marginal parameters (including marginal variance, shape, range, and tail decay parameters) can take arbitrary positive values.  The only condition is that $\bfA$ is a correlation matrix. Thus, this condition appears to be quite simple yet flexible for practical deployment compared to Corollary 3 in \citet{Apanasovich2012}, as many existing conditions impose inequality constraints. 
\end{remark}

\begin{remark}
Both models permit two kinds of asymmetries: space-time asymmetry and asymmetry between different processes at a fixed time lag. These asymmetric structures are illustrated via their contour plots given in Figures~\ref{fig: mvn multiple velocity} and~\ref{fig: mvn asymmetry} of Appendix~\ref{app: multivariate}. 
\end{remark}

\subsection{A General Form of Asymmetric Covariance Models}
%The proposed covariance functions in Equations~\eqref{eqn: LagMatern cov},~\eqref{eqn: LagCH cov}, and~\eqref{eqn: LagGH} are all based on the Lagrangian reference framework via location mixtures. 

An interesting question would be whether the Lagrangian reference framework would give analytical expressions for any choice of isotropic correlation functions.  As noted in \cite{Schlather2010}, analytical expressions are generally not available for arbitrary choice of isotropic correlations. Thanks to the proposed hierarchical mixture approach, closed-forms are possible to obtain. What follows is to consider a general form of asymmetric covariance functions. 

Let $\varphi(\cdot)$ be an isotropic correlation that is positive-definite in $\mathbb{R}^d$ for any fixed $d$. Define a real-valued function $\psi: \mathbb{R}^d \times \mathbb{R} \to \mathbb{R}$ such that
\begin{align} \label{eqn: general Lag form}
    \psi(\bfh, u) = \sigma^2 |\mathbf{I}_{d\times d} + u^2 \bfLambda|^{-a/(2d)} \varphi\biggl( \bigl\{(\bfh - \bflambda u)^\top (\bfI_{d\times d} + u^2 \bfLambda)^{-1} (\bfh - \bflambda u)\bigr\}^{1/2} \biggr),
\end{align}
where $\sigma^2>0, a>0, \bflambda \in \mathbb{R}^d$, $\bfLambda$ is a $d\times d$ positive definite matrix. The form~\eqref{eqn: general Lag form} is not directly derived based on the Lagrangian framework, since for $\bfv \sim \mathcal{N}(\bflambda, \bfLambda)$, $\psi(\bfh, u)\neq E_{\bfv} [\varphi(\|\bfh - \bfv u\|)]$ for any isotropic correlation $\varphi(\cdot)$ in general except for the Gaussian case. 
%However, the form~\eqref{eqn: general Lag form} includes the proposed covariance functions in Equations~\eqref{eqn: LagMatern cov},~\eqref{eqn: LagCH cov}, and~\eqref{eqn: LagGH} as special cases; 
Because its form coincides with the Lagrangian Gaussian covariance, one might assert that the function~\eqref{eqn: general Lag form} is  positive definite for any choice of isotropic correlation function $\varphi(\cdot)$. While this assertion has not been proven in the existing literature \citep[e.g.,][]{Schlather2010}, this assertion has been assumed in the existing works \citep{Salvana2023, Zhang2024, Salvana2020}.  The following result proves its positive definiteness, thus closing this gap.  

\begin{theorem} \label{thm: GLag framework}
Assume that $\varphi$ is a continuous isotropic correlation function. 
Then the function $\psi(\cdot, \cdot)$ in Equation~\eqref{eqn: general Lag form} is a positive-definite covariance function for any $a\geq d$. Moreover, For any fixed $\bfh$,  $\psi(\bfh, u) \sim \sigma^2 |\bfLambda|^{-a/(2d)} |u|^{-a} \varphi((\bflambda^\top \bfLambda^{-1}\bflambda)^{1/2})$ as $u \to \infty$. 
\end{theorem}

\begin{remark}
While the function~\eqref{eqn: general Lag form} is a valid space-time covariance function for any choice of isotropic correlation function $\varphi$, it is \textit{not} directly constructed from the Lagrangian framework. %As demonstrated in previous sections, the Lagrangian Mat\'ern model and the Lagrangian CH model could allow flexible origin behaviors but with possibly limited tail behavior in time. 
By specifying $\varphi$ as either the Mat\'ern correlation or the CH correlation, $\psi(\cdot)$ does not yield more flexible tail behaviors than Lagrangian Mat\'ern and CH models as $a\geq d$. 
Figure~\ref{fig: GL contours} of Appendix~\ref{app: general asymmetric models} shows their contour plots.       
\end{remark}

\begin{remark}
\cite{Salvana2023} construct their multivariate models based on the covariance function~\eqref{eqn: general Lag form} with the case $a=d$. Their results implicitly assume that the function~\eqref{eqn: general Lag form} is positive definite for any choice of $\varphi$ without a proof. The results in \cite{Salvana2023} are then restricted to the case where $\varphi$ is a squared-exponential correlation function \citep{Zhang2024}. The implication is  that the multivariate models in \cite{Salvana2023} have limited smoothness and tail behaviors when only considering a squared-exponential correlation function for $\varphi$. 
%While the claim that the function~\eqref{eqn: general Lag form} with $a=d$ is a positive definite covariance function has not been proven in \cite{Salvana2023} and prior literature, our result in Theorem~\ref{thm: GLag framework}  suggests that  $\psi(\cdot, \cdot)$ in~\eqref{eqn: general Lag form} is indeed positive definite for any choice of isotropic correlation function $\varphi$; thus Theorem 2 of \cite{Salvana2023} still holds true when combined with Theorem~\ref{thm: GLag framework} of this paper. %The resulting construction for multivariate space-time models is still valid. 
\end{remark}

%\begin{remark}
%For the case where $a=d$ and $\varphi$ is taken to be a squared-exponential (or Gaussian) correlation function, the resulting space-time covariance function becomes a Lagrangian Gaussian covariance~\eqref{eqn: LagGauss Cov}; see also Example 9 of \citet{Schlather2010}.  %\citetext{see also Example 9 of \citealt{Schlather2010}}. %The positive definiteness follows from the proof in Corollary 1.3 of \citet{Ma2003JSPI} as it is derived from the Lagrangian reference framework.
%\end{remark}

\section{Hierarchical Mixtures for Existing Covariance Models} \label{sec: hierarchical mixtures of existing models}
This section illustrates how this hierarchical mixture  approach can recover several existing models. Some specific examples are given below and in Appendix~\ref{app: existing models}. 

\begin{example}[Gneiting's class] \label{ex: Gneiting}
\cite{Gneiting2002} proposes a general class of nonseparable space-time covariance functions that are fully symmetric with the form:
\begin{align*} %\label{eqn: Gneiting cov}
    C(\mathbf{h}, u) = \frac{\sigma^2}{\psi(|u|^2)^{d/2}} \varphi\left( \frac{\|\mathbf{h}\|^2}{\psi(|u|^2)} \right), \quad (\mathbf{h}, u) \in \mathbb{R}^d \times \mathbb{R}
\end{align*}
where $\varphi(t), t\geq 0$ is a completely monotone function, and $\psi(t),t\geq 0$ is a positive function with a completely monotone derivative. This covariance class contains many examples of the covariance functions in \cite{Cressie1999}. As shown in Appendix~\ref{app: Gneiting}, the covariance class can be represented as a hierarchical mixture form. 
\end{example}

\begin{example} \label{ex: stein}
Consider the following reparameterized space-time covariance function in \citet[][Eq.~(12), p.~315]{Stein2005} with the Markov property:
\begin{align*}
C(\mathbf{h}, u) = \sigma^2 \frac{\Gamma(\nu+d/2) \Gamma(\nu+\zeta |u|)}{\Gamma(\nu) \Gamma(\nu+\zeta |u| + d/2)}  \mathcal{M}(\| \bfh - \epsilon u \bfe\|; \nu + \zeta|u|, \phi),\quad (\mathbf{h}, u) \in \mathbb{R}^d \times \mathbb{R}
\end{align*}
where $\sigma^2>0, \nu>0, \zeta>0, \phi>0, \epsilon>0$. $\bfe$ is a unit vector. As shown in Appendix~\ref{app: stein}, this covariance has a hierarchical mixture form. \citet[][p.~315]{Stein2005} points out that this covariance function only allows  algebraic decay in time along a certain axis at a fixed rate of decay, thus limiting its flexibility in capturing long-range dependence in time. Moreover, this asymmetry structure is an example of the frozen field Lagrangian framework. These limitations could  be addressed via the proposed hierarchical mixture approach. 
\end{example}

There are other constructions of space-time models with closed-form expressions using discrete mixtures \citep[e.g.,][]{Ma2002Geo, Ma2003JSPI, Porcu2007, Porcu2008, DeIaco2002}, but the resultant covariance models are generally difficult to work with in practice since parameter estimation in these models is challenging within the likelihood-based framework including the maximum likelihood estimation and Bayesian estimation. Thus, these covariance models are omitted from the discussion here.

%\vspace{-1cm}

\section{Applications} \label{sec: numerical example}
\subsection{Irish Wind Data} \label{sec: irish wind}
The classic Irish wind dataset originally analyzed by \cite{Haslett1989} provides daily observations over 18 years at 12 stations without missing data. The dataset is available in the \textsf{R} package \texttt{gstat} and has become a benchmark dataset for testing the performance of stationary space-time covariance models \citep[][]{Stein2005, Gneiting1997, Horrell2017, Stein2005JRSSB}. Following the same approach in \cite{Haslett1989} and \cite{Gneiting2002}, we fit various space-time covariance models to the square root of the wind speed after removing the seasonal trend and one of the sites, Rosslare, due to its clear nonstationarity. The analysis is based on  $n=72,314$ observations. 

This dataset presents several challenges for the analysis. First, it is computationally infeasible to evaluate the full likelihood, and hence approximate likelihood approaches \citep[e.g.,][]{Stein2005, Horrell2017} are often used. Here we consider the grouped Vecchia approximation \citep{Guinness2018, Katzfuss2021}, which extends the original Vecchia approximation \citep{Vecchia1988} with better approximation accuracy. Second, as suggested in previous works, this dataset exhibits long memory dependence in time and has an asymmetric dependence structure between space and time. Thus, it is desirable to use models with algebraic decay in time and an asymmetric structure such as the Lagrangian CH model~\eqref{eqn: LagCH cov}. While previous works have suggested various ways to compare models including semivariograms and likelihood, we shall focus on the likelihood-based comparison. We also add an overall nugget effect to the model of the form $\tilde{K}(\bfh, t)=  K(\bfh, t) + \sigma^2 \tau^2 \mathbb{I}(\bfh=0, t=0)$, where $K$ is a space-time covariance model with variance parameter $\sigma^2$ and $\tau^2$ denotes the noise to signal ratio. A more flexible nugget to account for the purely spatial and purely temporal nugget can also be used \citep{Stein2005}.  All the model comparison results are shown in Table~\ref{table: irish wind} for the Lagrangian Gaussian  model~\eqref{eqn: LagGauss Cov}, the Lagrangian Mat\'ern  model~\eqref{eqn: LagMatern cov}, and the Lagrangian CH model~\eqref{eqn: LagCH cov}. 
\begin{table}[!ht]
\centering
\normalsize
   \caption{Model comparison for Irish wind data under different covariance models. AIC represents Akaike information criterion and BIC represents Bayesian information criterion. The grouped Vecchia approximation is used for likelihood evaluation with different numbers of neighbors $m$.}
  {\resizebox{1.0\textwidth}{!}{%
  \setlength{\tabcolsep}{2.0em}
   \begin{tabular}{l c c c c} 
   \toprule \noalign{\vskip 1.5pt} 
Model & log-likelihood & AIC & BIC & \# of cov. parameters \\ \noalign{\vskip 1.5pt} 
\midrule \noalign{\vskip 1.5pt} 
%\multicolumn{5}{c}{ \ul{$m=50$}  }\\  \noalign{\vskip 3.5pt} 
%L-Gauss  & -29,813 & 59,642 & 59,716 & 8\\ \noalign{\vskip 2.5pt} 
%L-Mat\'ern  & -13,115 & 26,249 & 26,331 & 9\\ \noalign{\vskip 2.5pt} 
%L-CH   & -13,114 & 26,249 & 26,341 & 10 \\ \noalign{\vskip 2.5pt} 

%\midrule \noalign{\vskip 2.5pt} 
\multicolumn{5}{c}{\ul{$m=100$} } \\  \noalign{\vskip 3.0pt} 
Lagrangian Gaussian  & -28,843 & 57,703 & 57,716 & 8\\ \noalign{\vskip 2.5pt} 
Lagrangian Mat\'ern  & -11,930 & 23,878 & 23,961 & 9\\ \noalign{\vskip 2.5pt} 
Lagrangian CH   & -11,944 & 23,908 & 24,000  & 10 \\ \noalign{\vskip 2.5pt} 
\multicolumn{5}{c}{\ul{$m=150$} } \\  \noalign{\vskip 2.5pt} 

Lagrangian Gaussian  & -28,402 & 56,821 & 56,894 & 8\\ \noalign{\vskip 2.5pt} 
Lagrangian Mat\'ern  & -11,448 & 22,915 & 22,997 & 9\\ \noalign{\vskip 2.5pt} 
Lagrangian CH   & -11,419 & 22,858 & 22,950  & 10 \\ \noalign{\vskip 2.5pt} 
%GL-Mat\'ern & -8,577  & 17,174  & 17,266 &  10 \\ \noalign{\vskip 2.5pt} %file: 8452597.out 
\bottomrule
   \end{tabular}%
   }}
   \label{table: irish wind}
\end{table}  

All the parameters are estimated using the grouped Vecchia approximation with sufficiently large number of neighbors selected via maximin design in the \textsf{R} package \texttt{GpGp}. After extensive experimentation, the best fitted results are reported when the algorithms did not show lack of convergence. The results in Table~\ref{table: irish wind} indicate that  Lagrangian Mat\'ern and  CH models fit much better than the existing Lagrangian Gaussian model  in terms of log-likelihood, AIC and BIC, meaning that allowing flexible smoothness behavior can improve the model fit substantially. Increasing the number of neighbors in grouped Vecchia approximation from 50 (Table~\ref{table: irish wind m=50} of Appendix~\ref{app: numerical results}) to 150 improves model fit, but further improvements in likelihood  become negligible with more than 150 neighbors when taking into  computing time in consideration. An interesting finding is that the Lagrangian CH model gives a slightly better likelihood than Lagrangian Mat\'ern with  150 neighbors, which indicates that the spatial correlation exhibits faster decay than the temporal decay and the Lagrangian CH model can correctly fall back to the behavior in the Lagrangian Mat\'ern model. This is expected as the Lagrangian Mat\'ern model can be viewed as a special case of the Lagrangian CH model. While existing works \citep{Haslett1989, Gneiting2002} suggested long-range dependence in time, this result above suggests that a covariance model with algebraic decay suffices to capture the slow decay behavior in time, which also agrees with the finding in \cite{Horrell2017}.

The parameter estimates for all these models are presented in Table~\ref{table: irish wind param} of Appendix~\ref{app: numerical results}. For  Lagrangian Mat\'ern and CH models, their smoothness parameters are estimated to be 0.41, indicating a mean-square continuous but not differentiable space-time process. The estimates for Lagrangian Mat\'ern  and CH models share similar results in terms of smoothness parameter $\nu$, nugget parameter $\tau^2$, variance parameter $\sigma^2$, and other asymmetry parameters $\lambda_0, \theta_0, \theta_1, \lambda_1, \lambda_2$. The tail decay parameter $\alpha$ in the Lagrangian CH model is estimated to be 3.44, indicating very fast algebraic decay in space and justifying why   Lagrangian Mat\'ern and CH models give very similar likelihoods.  Compared with the best fitted models in \citet[][p.~318]{Stein2005}, both Lagrangian Mat\'ern  and CH models  give much smaller nugget estimates, even though \cite{Stein2005} considered a flexible nugget term that accounts for distinct nugget effects in space and in time. This  implies that Lagrangian Mat\'ern  and CH models fit the data much better than the asymmetric covariance model in \cite{Stein2005}, partly because the two best fitted models in \cite{Stein2005} correspond to frozen fields while the proposed models have a more flexible asymmetric structure; and partly because Lagrangian Mat\'ern  and  CH models allow algebraic decay in time  while the best fitted models in \cite{Stein2005} with asymmetry only exhibit algebraic decay along certain directions in space-time domain.  Another comparison with the half-spectral model \citep[][Table 3]{Horrell2017} suggests that Lagrangian Mat\'ern  and CH models give much better results than the fitted asymmetric model with temporal tail decay in \cite{Horrell2017}, even though our results were fitted via the grouped Vecchia likelihood compared to the full likelihood evaluation \citep{Horrell2017}.  

\subsection{Air Temperature Data}
Air temperature (AT) is a key climate variable in understanding Earth systems for many geophysical processes and their interactions. Air temperature has been a primary driver for studying urban heat island (UHI) effects due to rapid urbanization and socioeconomic development \citep{Estrada2017, Li2017}. Monitoring observations and climate proxies such as remote sensing data are widely available. For instance, Global Historical Climatology Network-Daily (GHCNd) is a database that includes maximum and minimum daily temperature over global land areas for climate analysis and monitoring studies, and is managed by the National Oceanic and Atmospheric Administration (NOAA) National Centers for Environmental Information (NCEI) \citep{Menne2012b}. This database contains a long historical record with high resolution, but the observations are irregularly monitored and sparse in space. Statistical analysis is often performed to produce  high-resolution air temperature data with complete spatio-temporal coverage by synthesizing information from other climate proxies \citep{Li2018a, Li2018b}. 

This example considers the daily air temperature dataset at 689 weather stations in California  in the year 2010. This study region involves many complex geophysical processes due to many factors including land-sea interaction, environmental profiles, and socioeconomic factors. There are $190,643$ AT observations in California in 2010. While air temperature  exhibits complex space-time structures, its variability could be largely explained by land surface temperature (LST) as they appear to be highly correlated \citep{Li2018a, Li2018b}. Following \cite{Li2018b}, we use the gap-filled MODIS daily LST data produced based on MODIS LST daily observations, MOD11A1 and MYD11A1 from Terra and Aqua satellites, respectively \citep{Li2018a}. The LST dataset matches the AT dataset in space-time locations in the year 2010 except on June 11 and 12, 2010, when LST data were missing.  The elevation data is also used as an important covariate. The detailed pre-processing procedure for both LST and elevation data is given by \cite{Li2018b}. 

\begin{figure}[!ht]
\begin{center}
\makebox[\textwidth][c]{\includegraphics[width=1\textwidth, height=0.5\textheight]{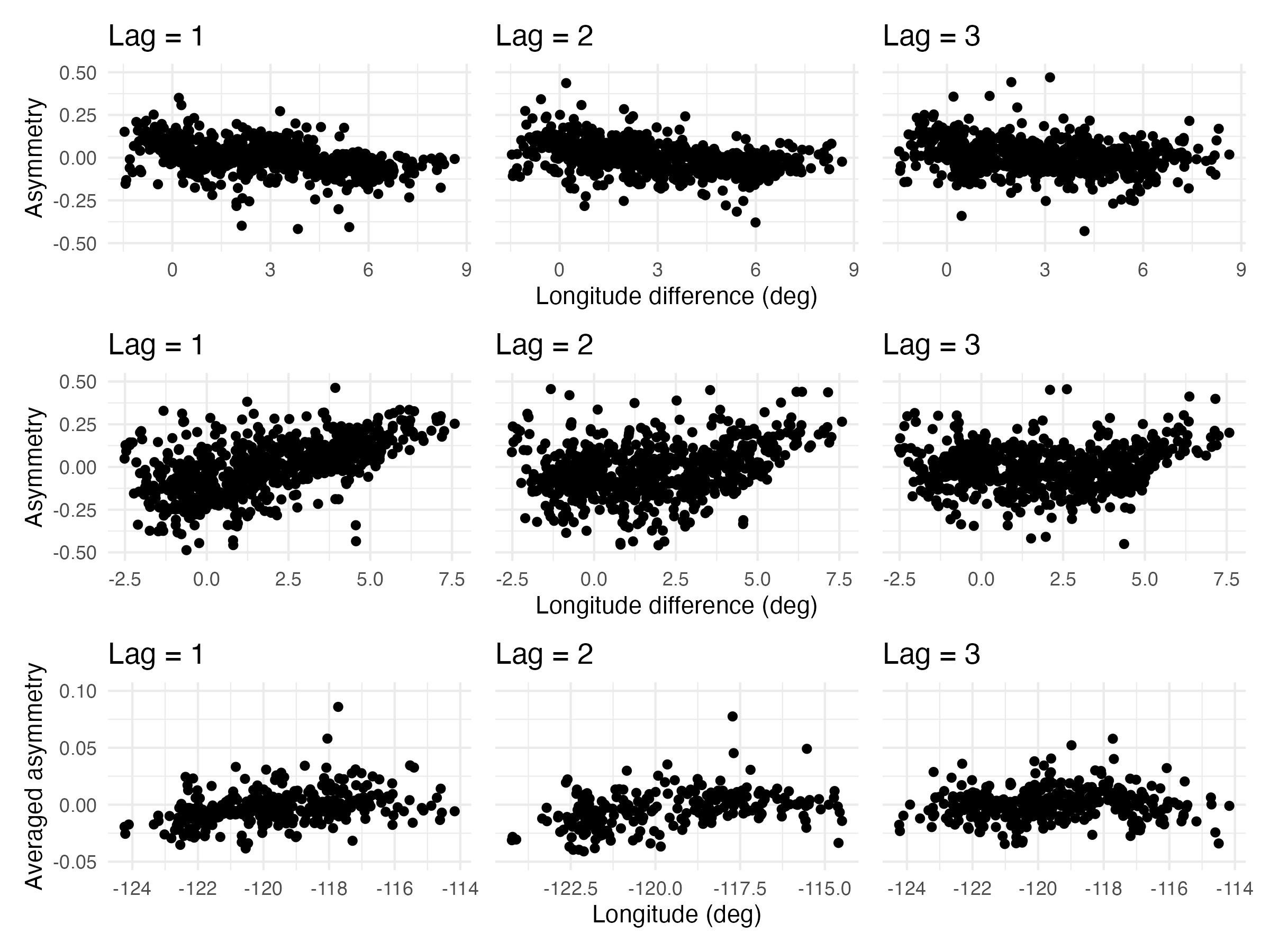}}
\caption{Empirical space-time semivariogram differences at temporal lags $k=1, 2, 3$. Both the first  and second rows show $\delta(\bfs_j, \bfs_1; k)$ on the vertical axis against longitudinal difference in $\bfs_j - \bfs_1$ over all locations $j=1,\ldots, n_s$. In the first row, $\bfs_1=(-122.80, 41.31)$ which is in the northwest of the location $\bfs_1=(-121.75, 37.98)$ in the second row. The third row shows $\bar{\delta}(\bfs_j; k)$ across all locations $j=1,\ldots, n_s$.} 
\label{fig: AT variograms}
\end{center}
\end{figure}

To simplify the analysis, we first perform exploratory analysis to remove seasonal effects and the covariates including LST, elevation, longitude, and latitude. We then scale the residuals by standard deviation on each day to obtain a more homogeneous space-time dataset, which is denoted by $z(\cdot, \cdot)$. Exploratory data analysis on the residuals $z(\cdot, \cdot)$ shows a clear lack of nonstationary features in space and in time. To check  space-time asymmetric dependence structures, we define the empirical space-time semivariogram of space-time process $\{z(\bfs,t): (\bfs, t) \in \mathbb{R}^d\times \mathbb{R}\}$ at temporal lag $k\geq 0$, 
\begin{align*}
    g(\bfs_1, \bfs_2; k) = \frac{1}{2(n_t-k)} \sum_{j=1}^{n_t - k} \{z(\bfs_1, j+k) - z(\bfs_2, j) \}^2, 
\end{align*}
where $n_t$ denotes the number of observed time points. We then define $\delta(\bfs_1, \bfs_2; k) := g(\bfs_1, \bfs_2; k) - g(\bfs_2, \bfs_1; k)$ to be the empirical space-time semivariogram difference between locations $\bfs_1$ and $\bfs_2$ at lag $k$. If $z(\cdot, \cdot)$ is stationary in space-time with covariance function $K$, $E\{\delta(\bfs_1, \bfs_2; k) \} = K(\bfs_2 - \bfs_1, k) - K(\bfs_1-\bfs_2; k)$. If $K$ is fully symmetric, then $\delta$ would be randomly scattered around zero for all locations and all temporal lags. The lack of a zero average pattern in $\delta$ necessarily indicates that $K$ is asymmetric. We further define $\bar{\delta}(\bfs; k) := {1}/{n_s} \sum_{i=1}^{n_s} \delta(\bfs_i, \bfs; k) $ to be the average of empirical space-time semivariogram differences over $n_s$ locations given location $\bfs$ and lag $k$. If $\bar{\delta}$ shows non-zero patterns on average across all locations, $K$ is necessarily asymmetric. This function can show the averaged empirical asymmetric structure for any given set of observation locations. Figure~\ref{fig: AT variograms} shows clear asymmetry structures and such asymmetry structures also vary across temporal lags, implying that a frozen-field Lagrangian reference covariance model could be undesirable. 

To test the performance of asymmetric covariance models, we model $z(\cdot, \cdot)$ as a stationary space-time process with unknown constant mean and several proposed covariance models. The best fitted results  reported in Table~\ref{table: AT} indicate that Lagrangian CH and Mat\'ern fit much better than Lagrangian Gauss model while the Lagrangian CH model fits better than the Lagrangian Mat\'ern model in terms of likelihood, AIC, and BIC. This indicates that allowing both flexible smoothness and algebraic tails is desirable. It appears that using a large number of neighbors can considerably improve model fit via grouped Vecchia approximation. The parameter estimates in Table~\ref{table: AT param} of Appendix~\ref{app: numerical results} show that Lagrangian Mat\'ern  and CH models give similar parameter estimates in terms of $\nu, \tau^2, \sigma^2, \lambda_0, \theta_0, \theta_1, \lambda_1, \lambda_2$. As expected, the tail decay parameter $\alpha$ in Lagrangian CH model is large (around 3.1), indicating very fast decay in space. This partly explains why the Lagrangian Mat\'ern model could give very close model fitting results. 
\begin{table}[!ht]
\centering
\normalsize
   \caption{Model comparison for air temperature data. AIC represents Akaike information criterion and BIC represents Bayesian information criterion. The grouped Vecchia approximation is used for likelihood evaluation with different numbers of neighbors $m$.}
  {\resizebox{1.0\textwidth}{!}{%
  \setlength{\tabcolsep}{2.0em}
   \begin{tabular}{l c c c c} 
   \toprule \noalign{\vskip 1.5pt} 
Model & log-likelihood & AIC & BIC & \# of cov. parameters \\ \noalign{\vskip 1.5pt} 
\midrule \noalign{\vskip 1.5pt} 
%\midrule \noalign{\vskip 2.5pt} 
\multicolumn{5}{c}{\ul{$m=100$} } \\  \noalign{\vskip 3.0pt} 
Lagrangian Gaussian & -235,419 & 470,854 & 470,935 & 8 \\ \noalign{\vskip 2.5pt}
%L-Mat\'ern  & -228,192 & 456,223 & 456,314 & 9\\ \noalign{\vskip 2.5pt} % non-scaled
Lagrangian Mat\'ern  & -194,541 & 389,101 & 389,192 & 9\\ \noalign{\vskip 2.5pt} % scaled
%L-CH   & -228,183  & 456,386  & 456,488   & 10 \\ \noalign{\vskip 2.5pt} 
Lagrangian CH   & -194,490  & 389,000  & 389,101   & 10 \\ \noalign{\vskip 2.5pt} % scaled
\multicolumn{5}{c}{\ul{$m=150$} } \\  \noalign{\vskip 2.5pt} 

Lagrangian Gaussian & -232,351  & 464,718 & 464,799 & 8 \\  % scaled
%L-Mat\'ern  &-219,791  & 439,601  & 439,692 & 9\\ \noalign{\vskip 2.5pt} % non-scaled
Lagrangian Mat\'ern  &-190,872  & 381,762  & 381,853 & 9\\ \noalign{\vskip 2.5pt} % scaled file: 8688388.out
Lagrangian CH   &-190,686  & 381,392  & 381,494  & 10 \\ \noalign{\vskip 2.5pt}   % scaled filed: 8532147.out
\bottomrule
   \end{tabular}%
   }}
   \label{table: AT}
\end{table} 
%\newpage

\section{Discussion} \label{sec: discussion}

This paper introduces a hierarchical mixture framework that not only constructs a new class of asymmetric space-time covariance functions but also provides a direct connection between these covariance models and the associated space-time processes. This framework is a constructive approach that can produce asymmetric space-time covariance functions and allows extensions for nonstationary and multivariate models with closed forms. The framework explicitly integrates the Lagrangian framework for physical interpretation and provides Lagrangian Mat\'ern and CH covariance models, and thus bridges the gap in the existing literature \citep{Cox1988, Schlather2010, Salvana2023} as the only existing option is the Lagrangian Gaussian model, which is infinitely differentiable and has exponential tails in space and in time.  Several key aspects have been studied in terms of analytic tractability, practical usefulness, and model assessment. The practical usefulness was examined in terms of smoothness near the origin, long-range dependence, and space-time asymmetry, which are clearly not an exhaustive list of theoretical bases for modeling complex space-time processes.

There are several potential improvements that could be further explored in the current work. The current work constructs the covariance functions in $\mathbb{R}^d\times \mathbb{R}$. For  space-time data over a large geographical region, it might be desirable to directly construct these models over sphere by applying techniques in \cite{Porcu2016}. The numerical experiments indicated that the grouped Vecchia likelihood \citep{Guinness2018, Katzfuss2021} can gain considerable improvements using large number neighbors in space-time domain at the cost of more computing time. This computational challenge and the loss of accuracy via Vecchia approximate likelihood compared with the full likelihood deserve more careful investigation in the space-time setting as space-time datasets are often much larger and parameter estimation is more expensive due to the inclusion of more parameters. The numerical examples indicated that the log likelihood could be maximized for certain parameters on the boundary of the parameter space, which is likely due to the flatness of (log) likelihood for certain parameters. This issue suggests a close analog of the same issue in the likelihood functions for spatial processes \citetext{\citealp[p.~173]{Stein1999}, \citealp{Berger2001, Zhang2004, Ma2023}}. It would be interesting to  elicit prior distributions for model parameters to achieve robust estimation of parameters.  

This paper is mainly focused on investigating the theoretical properties of this hierarchical specification. The proposed hierarchical mixture framework presents several research directions that could be further explored. First, a promising direction that has not been fully explored is to integrate hierarchical construction of space-time processes with statistical inference from a hierarchical modeling perspective, as opposed to the standard practice of treating covariance model construction and its inference separately. For instance, if one has reliable flow information regarding asymmetry at desirable spatio-temporal scales, such information could be exploited to elicit the (prior) probability distribution of the velocity vector \citep{Huang2004} in the hierarchical mixture framework. Second, it is possible to use this hierarchical mixture approach for modeling non-Gaussian data \citep{FONSECA2011biometrika,Huser2019} and network data \citep{Tang2024}. Third, the hierarchical mixture construction of asymmetric space-time processes also has a close connection to physical partial differential models. Techniques such as inferential strategies and computations for Mat\'ern-based stochastic partial differential equations would  be useful in statistical inference for the physical equations corresponding to the proposed covariance models. Finally, the existing hierarchical mixture in nonstationarity and multivariate extensions can be further explored to allow increased flexibility \citep{Ma2026EJS}.  

 \vspace{-.2cm} 
\section*{Supplementary Materials}

The Supplementary Materials contain an appendix (.pdf file) and  an \textsf{R} package with related \textsf{R} code to reproduce the numerical results (.zip file).

%\section*{Ackknowledgements}
%The author would like to thank Zhengyuan Zhu for suggesting the air temperature data and helpful comments on the draft of the manuscript.  

\bibliographystyle{biom}
\begin{singlespacing}
\bibliography{main}
\end{singlespacing}

\newpage

\section*{Appendix}
\appendix
\makeatletter 
% "activate" the preparatory code, but for section-level headers only
\newcommand{\section@cntformat}{Appendix \thesection:\ }
\newcommand{\subsection@cntformat}{Appendix \thesubsection:\ }
\makeatother
%\end{comment}

%%%% Main text entry area:
\setcounter{equation}{0}
\setcounter{page}{1}
\setcounter{table}{0}
\setcounter{section}{0}
\setcounter{figure}{0}
\numberwithin{table}{section}

\renewcommand{\theremark}{S.\arabic{remark}}
\renewcommand{\theequation}{S.\arabic{equation}}

\renewcommand{\thepage}{S.\arabic{page}}
\renewcommand{\thetable}{S.\arabic{table}}
\renewcommand{\thefigure}{S.\arabic{figure}}
\renewcommand{\theproposition}{S.\arabic{proposition}}
\renewcommand{\theexample}{S.\arabic{example}}

\baselineskip=25pt

%\clearpage\pagebreak\newpage

\spacingset{1.65} % DON'T change the spacing!

%\onehalfspacing 

This supplement contains three sections. Section~\ref{sec: proofs} contains all the technical proofs for the main results. Section~\ref{sec: mixture construction} contains mixture constructions for  new space-time covariance models and several existing space-time covariance models. Section~\ref{app: numerical results} contains additional numerical results. 

\section{Technical Proofs} \label{sec: proofs}

\subsection{Proof of Theorem~\ref{thm: pd}} \label{app: pd}
\begin{proof}
Note that for all $\Ell \in \mathbb{R}^d\times \mathbb{R}$, $C(\Ell, \Ell) = C_s(\mathbf{0}) \int_{\mathbb{R}^d}  f^2(\bfv|\bfxi_{\Ell}) \mu(\mathrm{d} \bfv) <\infty$ since $f(\cdot|\bfxi_{\Ell}) \in L^2(\mu)$. By assumption, it remains to check positive-definiteness for both $C$ and $K$. 

We first prove the positive-definiteness of $C$. For any positive integer $n$, arbitrary real numbers $a_1, \ldots, a_n \in \mathbb{R}$, and any set $\{\Ell_i=(\bfs_i, u_i) \in \mathbb{R}^d\times \mathbb{R}: i=1,\ldots, n\}$, let $\bfh_{ij}:=\bfs_i - \bfs_j$ and $u_{ij}:=u_i - u_j$. Then we have 
\begin{align*}
    \sum_{i=1}^n  \sum_{j=1}^n a_i a_j C(\Ell_i, \Ell_j) &= \sum_{i=1}^n \sum_{j=1}^n a_i a_j \int_{\mathbb{R}^d} C_s(\bfh_{ij}-\bfv u_{ij}) f(\bfv|\bfxi_{\Ell_i}) f(\bfv|\bfxi_{\Ell_j}) \mu(\mathrm{d}\bfv) \\
    &= \int_{\mathbb{R}^d} \sum_{i=1}^n \sum_{j=1}^n (a_i f(\bfv|\bfxi_{\Ell_i})) C_s(\bfh_{ij} - \bfv u_{ij}) (a_j f(\bfv|\bfxi_{\Ell_j})) \mu(\mathrm{d} \bfv) \\
    &= \int_{\mathbb{R}^d} (\mathbf{a} \circ \mathbf{f})^\top \mathbf{C}_s (\mathbf{a} \circ \mathbf{f}) \mu(\mathrm{d} \bfv) \geq 0
\end{align*}
where $\mathbf{a}:=(a_1,\ldots, a_n)^\top$ and $\mathbf{f}:=[f(\bfv| \bfxi_{\Ell_1}), \ldots, f(\bfv| \bfxi_{\Ell_n})]^\top$ are both $n$-dimensional vectors and $\mathbf{C}_s:=[C_s(\bfh_{ij}-\bfv u_{ij})]_{i,j=1,\ldots, n}$ is a positive definite matrix since $C_s$ is positive definite in $\mathbb{R}^d$ by assumption. $\circ$ denotes elementwise multiplication. The equality holds if and only if $a_i=0, \forall i=1,\ldots, n$. Therefore $C$ is positive-definite.  Similarly, we observe that 
\begin{align*}
    \sum_{i=1}^n \sum_{j=1}^n a_i a_j K(\Ell_i, \Ell_j) &= \sum_{i=1}^n \sum_{j=1}^n a_i a_j \int_{\mathbb{R}^p} C(\Ell_i, \Ell_j; \bfzeta) g(\bfeta| \bflambda_{\Ell_i}) g(\bfeta| \bflambda_{\Ell_j}) \nu(\mathrm{d} \bfeta) \\
   &= \int_{\mathbb{R}^p} \sum_{i=1}^n \sum_{j=1}^n (a_i g(\bfeta| \bflambda_{\Ell_i})) C(\Ell_i, \Ell_j; \bfzeta) (a_j g(\bfeta| \bflambda_{\Ell_j})) \nu( \mathrm{d} \bfeta) \\
    & = \int_{\mathbb{R}^p} (\mathbf{a}\circ \mathbf{g})^\top \mathbf{C} (\mathbf{a}\circ \mathbf{g}) \nu(\mathrm{d} \bfeta )  \geq 0
\end{align*}
where $\mathbf{g}:=[g(\bfeta| \bflambda_{\Ell_1}), \ldots, g(\bfeta| \bflambda_{\Ell_n})]^\top$ is an $n$-dimensional vector. $\mathbf{C}:=[C(\Ell_i, \Ell_j)]_{i,j=1,\ldots, n}$ is a positive-definite matrix from previous arguments. Thus, $K$ is positive semidefinite. %The equality holds if and only if $a_i=0, \forall i=1,\ldots, n$. Thus $K$ is also positive-definite.

\end{proof}

\subsection{Proof of Theorem~\ref{thm: tail behavior}} \label{app: tail behavior}
\begin{proof}
Assume that $R$ is well-defined. To prove (i), we take the Taylor expansion of the Gaussian function $\exp(-\|\bfh\|^2/(2U))$ for $\|\bfh\|$ near 0, and obtain 
\begin{align*}
R(\bfh) = \int_0^{\infty} \sum_{k=0}^{\infty} \frac{(-1)^k \|\bfh\|^{2k}}{k! 2^k U^k}  \mathrm{d} G(U)  
 =  \sum_{k=0}^{\infty}  \frac{(-1)^k }{k! 2^k} \biggr\{\int_0^{\infty} U^{-k} \mathrm{d} G(U)\biggr\} \|\bfh\|^{2k}. 
\end{align*}
It follows from Theorem 2 of \citet[][p.~29]{Stein1999} that if the expectation $E(U^{-k})$ exists, then $R$ has $2k$ derivatives and hence the corresponding Gaussian process is $k$-times mean-square differentiable. 

To prove (ii), we first write the integral in Gaussian mixture form as
\begin{align*}
R(\bfh) = \int_0^{\infty}  (2\pi U)^{-1/2}  \exp\left\{ - \frac{\|\bfh\|^2}{2U} \right\}  (2\pi U)^{1/2} g(U) \mathrm{d} U
\end{align*} 
Applications of Theorem 5.2 and Theorem 6.1 of \citet{Barndorff1982} yield the desired results.

\end{proof} 

\subsection{Proof of Proposition~\ref{prop: LagMatern}} \label{app: LMatern}
\begin{proof}
According to Equation~\eqref{eqn: LagGauss Cov} or \citet[][p.~794]{Schlather2010}, 
\begin{align} \nonumber
    C(\mathbf{h}, u; \rho^2) &= E_{\bfv}[C_s(\|\bfh - \bfv u\|)] \\ \label{eqn: Gaussian stcov}
    & = \sigma^2 | \mathbf{I}_{d\times d} + u^2 \bfLambda|^{-1/2} \exp \left\{ - \frac{(\mathbf{h}-u\boldsymbol \lambda)^\top (u^2 \bfLambda + \mathbf{I}_{d\times d})^{-1} (\mathbf{h}-u\boldsymbol \lambda) }{\rho^2} \right\}.
\end{align}
Let $r_u^2:=(\mathbf{h}-u\boldsymbol \lambda)^\top (u^2 \bfLambda + \mathbf{I}_{d\times d})^{-1} (\mathbf{h}-u\boldsymbol \lambda)$. The integral representation of the modified Bessel function of the second kind is $\mathcal{K}_{\nu}(z) = \frac{z^\nu}{2^{\nu+1}} \int_0^{\infty} t^{-\nu -1} \exp\{-t - z^2/(4t)\} \text{d} t$. As $\rho^2 \sim \text{Ga}(\nu, 1/(4\phi^2))$, it follows from direct computation and Equation~\eqref{eqn: Gaussian stcov} that 
\begin{align} \nonumber
    K(\bfh, u) &= \int C(\mathbf{h}, u; \rho^2) \text{Ga}(\rho^2| \nu, 1/(4\phi^2)) \, \text{d} \rho^2 \\  \nonumber
    &= \sigma^2 |\bfI_{d\times d} + u^2 \bfLambda|^{-1/2} \frac{(4\phi^2)^{-\nu}}{\Gamma(\nu)} \int  (\rho^2)^{\nu-1} \exp\left\{ - \frac{r_u^2}{\rho^2} -  \frac{\rho^2}{4\phi^2}  \right\} \mathrm{d}\rho^2 \\ \label{eqn: integral LagMatern}
    & \overset{x=r_u^2/\rho^2}{=\joinrel=\joinrel=\joinrel=} \sigma^2 |\bfI_{d\times d} + u^2 \bfLambda|^{-1/2} \frac{(4\phi^2)^{-\nu}}{\Gamma(\nu)} \int_0^{\infty} (r_u^2)^{\nu} x^{-\nu-1} \exp\{- x - r^2_u  /(4\phi^2x) \} \text{d} x \\ \nonumber
    &=\sigma^2 |\bfI_{d\times d} + u^2 \bfLambda|^{-1/2} \frac{(4\phi^2)^{-\nu}}{\Gamma(\nu)} (r_u^2)^{\nu} (r_u/\phi)^{-\nu} 2^{\nu+1} \mathcal{K}_{\nu}(r_u / \phi) \\  \nonumber
    &= \sigma^2 |\bfI_{d\times d} + u^2 \bfLambda|^{-1/2} \frac{2^{1-\nu}}{\Gamma(\nu)} ( r_u/\phi)^{\nu} \mathcal{K}_{\nu}(r_u/\phi) \\ \nonumber
    & = \sigma^2 |\bfI_{d\times d} + u^2 \bfLambda|^{-1/2} \mathcal{M}(r_u; \nu, \phi),
\end{align}
as desired. 
\end{proof}

\subsection{Proof of Proposition~\ref{prop: LagCH}} \label{app: LCH}
\begin{proof}
Based on the integral representation~\eqref{eqn: integral LagMatern}, we further mix over $\phi^2 \sim \text{IG}(\alpha, \beta^2/4)$. It follows from direct calculation that 
\begin{align*}
K(\bfh, u) %&=  \\
    &= \sigma^2 |\bfI_{d\times d} + u^2 \bfLambda|^{-1/2} \frac{1}{\Gamma(\nu) 2^{2\nu}} \\
    &\quad\times \int \int \biggl(\frac{r_u}{\phi} \biggr)^{2\nu} x^{-\nu-1} \exp\left\{-x - \frac{r_u^2 }{4\phi^2x} \right\} \frac{(\beta^2/4)^{\alpha}}{\Gamma(\alpha)} (\phi^2)^{-\alpha-1} \exp\left\{ -\beta^2 /(4\phi^2) \right\} \text{d}\phi^2 \text{d}x \\
    & = \sigma^2 |\bfI_{d\times d} + u^2 \bfLambda|^{-1/2} \frac{1}{\Gamma(\nu)\Gamma(\alpha) 2^{2\nu}} (\beta^2)^{\alpha} 2^{-2\alpha} \\ 
    &\quad \times \int r_u^{2\nu} x^{-\nu-1} \exp(-x) \biggr\{ \int_{0}^{\infty} (\phi^2)^{-(\nu+\alpha)-1} \exp\left\{ - \biggr[ r_u^2/(4x) + \beta^2/4 \biggr] / \phi^2 \right\} \text{d}\phi^2 \biggr\} \text{d}x \\
    & = \sigma^2 |\bfI_{d\times d} + u^2 \bfLambda|^{-1/2} \frac{1}{\Gamma(\nu)\Gamma(\alpha) 2^{2\nu}} (\beta^2)^{\alpha} 2^{-2\alpha} \\ 
    & \quad \times \int r_u^{2\nu} x^{-\nu-1} \exp(-x) \Gamma(\nu+\alpha) \biggr[ r_u^2/(4x) + \beta^2/4 \biggr]^{-(\nu+\alpha)} \text{d}x \\
    &\overset{x=r_u^2t/\beta^2}{=\joinrel=\joinrel=\joinrel=\joinrel=\joinrel=} \sigma^2 |\bfI_{d\times d} + u^2 \bfLambda|^{-1/2} \frac{\Gamma(\nu+\alpha)}{\Gamma(\nu)\Gamma(\alpha)} \int_{0}^{\infty} t^{\alpha-1}(1+t)^{-(\nu+\alpha)} \exp\left\{-\frac{r_u^2}{\beta^2} t \right\} \text{d}t \\
    &=\sigma^2 |\bfI_{d\times d} + u^2 \bfLambda|^{-1/2} \frac{\Gamma(\nu+\alpha)}{\Gamma(\nu)} \mathcal{U}\left(\alpha, 1-\nu, \frac{r_u^2}{\beta^2} \right) \\
    &= \sigma^2 |\bfI_{d\times d} + u^2 \bfLambda|^{-1/2} \mathcal{CH}\biggr( r_u; \nu, \alpha, \beta \biggr),
\end{align*}
where $\mathcal{U}(a,b,x):= \frac{1}{\Gamma(a)} \int_0^{\infty} t^{a-1} (t+1)^{-(1-b+a)} \exp(-{x}t) \text{d}t$ is the confluent hypergeometric function of the second kind \citep[][chap.~13.2]{Abramowitz1965}, and $\mathcal{CH}(x; \nu, \alpha, \beta) := \frac{\Gamma(\nu+\alpha)}{\Gamma(\nu)} \mathcal{U}(\alpha, 1-\nu, x^2/\beta^2)$ is the confluent hypergeometric correlation function. 

\end{proof}

\subsection{Proof of Proposition~\ref{prop: Lag nonstationarity}} \label{app: Lag nonstationary}
\begin{proof}
The covariance of $y(\cdot)$ given $\rho^2, \bfv$ is 
\begin{align*} 
C(\Ell_i, \Ell_j; \rho^2, \bfv):=\text{cov}(y(\Ell_i), y(\Ell_j)| \rho^2, \bfv) = \sigma^2 \exp\left\{ - \frac{ \| \bfh_{ij} - t_i \bfv(\Ell_i) + t_j \bfv(\Ell_j) \|^2 }{\rho^2} \right\},\,  \Ell_i, \Ell_j \in \mathbb{R}^{d+1},
\end{align*}
where $\bfh_{ij} = \bfs_i - \bfs_j$. 

By assumption, $\bfv(\cdot)$ is a $d$-variate Gaussian process conditional on $\rho^2$. It follows from the finite-dimensional distributions of $\bfv(\cdot)$ that 
\begin{align*}
t_i\bfv(\Ell_i) - t_j\bfv(\Ell_j) | \rho^2 \sim N_{d}\biggr (\bfm_{ij}, \frac{\rho^2}{2} \bfQ_{ij} \biggr)
\end{align*}
where 
$\bfm_{ij} := t_i \bflambda(\Ell_i) - t_j\bflambda(\Ell_j)$ and $\bfQ_{ij} :=t_i^2 \bfLambda(\Ell_i, \Ell_i) + t_j^2 \bfLambda(\Ell_j, \Ell_j) - t_i t_j [\bfLambda(\Ell_{i}, \Ell_j) + \bfLambda(\Ell_j, \Ell_i)]$.

Then mixing the process $y(\cdot)$ over the velocity field yields its covariance function 
\begin{align*}
C(\Ell_i, \Ell_j; \rho^2)&:=\text{cov}(y(\Ell_i), y(\Ell_j) | \rho^2) = E_{\bfv}\{\text{cov}(y(\Ell_i), y(\Ell_j) | \rho^2, \bfv) \} \\
& = \sigma^2 | \bfO_{ij} |^{-1/2} \exp \left( - \frac{r_{ij}^2}{\rho^2}\right)
\end{align*}
where $r_{ij}^2:=(\bfh_{ij} - \bfm_{ij})^\top \bfO_{ij}^{-1} (\bfh_{ij} - \bfm_{ij})$ and $\bfO_{ij}:=\bfI_{d} + \bfQ_{ij}$. The last equation follows by applying the same Gaussian integration identity as in Equation~\eqref{eqn: Gaussian stcov} in the proof of Proposition~\ref{prop: LagMatern}. 

To obtain a nonstationary Lagrangian Mat\'ern model, we could mix over $\rho^2 | \phi^2 \sim \text{Ga}(\nu,\allowbreak 1/(4\phi^2))$  and obtain 
\begin{align*}
\text{cov}(y(\Ell_i), y(\Ell_j) | \phi^2) &= \int C(\Ell_i, \Ell_j; \rho^2) \text{Ga}(\rho^2| \nu, 1/(4\phi^2))\mathrm{d}\rho^2 
=\sigma^2 | \bfO_{ij} |^{-1/2} \mathcal{M}(r_{ij}; \nu, \phi).
\end{align*}  

To obtain a nonstationary Lagrangian CH model, we could take a further mixture $\phi^2 \sim \text{IG}(\alpha, \beta^2/4)$. A direct calculation yields the nonstationary Lagrangian CH covariance function
\begin{align*}
\text{cov}(y(\Ell_i), y(\Ell_j)) &= E_{\phi^2}[\text{cov}(y(\Ell_i), y(\Ell_j) | \phi^2)] 
= \sigma^2 | \bfO_{ij} |^{-1/2} \mathcal{CH}(r_{ij}; \nu, \alpha, \beta).
\end{align*} 

\end{proof}

\subsection{Proof of Proposition~\ref{prop: mv mixture}}

\begin{proof} 
Let $\bfY(\cdot)$ denote a $q$-variate space-time process. 
The conditional cross-covariance of $\bfY(\cdot)$ given $\bfV, \rho^2$ is 
\begin{align*}
C_{ij}(\Ell_k, \Ell_m; \bfV, \rho^2) = \sigma_i \sigma_j a_{ij} \exp\left\{ - \frac{\| \bfs_k -\bfs_m - t_k\bfv_i  + t_m \bfv_j   \|^2}{\rho^2}  \right\}
\end{align*} 
which is positive definite since $[C_{ij}^0(\cdot, \cdot)]$ is positive definite by assumption. 

It follows from $\bfV | \rho^2 \sim \mathcal{N}_{qd}(\bfmu_{\lambda}, \bfSigma_{\lambda} \rho^2/2)$ that 
\begin{align*} 
t_k\bfv_i  - t_m \bfv_j  | \rho^2  \sim \mathcal{N}_{d}\bigg( t_k \bflambda_i - t_m \bflambda_j, \, \frac{\rho^2}{2} \bfQ_{ij} \bigg). 
\end{align*} 

Applying the same trick in Propositions 1 and 2 yields 
\begin{align*}
C_{ij}(\Ell_k, \Ell_m; \rho^2)=E_{\bfV}[C_{ij}(\Ell_k, \Ell_m; \bfV, \rho^2)] = \sigma_i \sigma_j a_{ij} |\bfO_{ij} |^{-1/2} \exp\left( - \frac{\bfh_{ij}^\top \bfO_{ij}^{-1} \bfh_{ij} }{\rho^2} \right)
\end{align*} 

It remains to derive the cross-covariance in parts (i) and (ii). Observe that 

\begin{align} \label{eqn: weight fun} \nonumber 
\sqrt{p_i(\rho^2| \phi^2) p_j(\rho^2|\phi^2)} &= \frac{ (1/(4\phi_i^2))^{\nu_i/2} (1/(4\phi_j^2))^{\nu_j/2} }{ \sqrt{\Gamma(\nu_i) \Gamma(\nu_j)}} (\rho^2)^{(\nu_i + \nu_j)/2-1} \exp\biggr\{-\frac{\rho^2[1/(4\phi_i^2) + 1/(4\phi_j^2)]}{2} \biggr\}  \\
&= c_{ij}^{\nu}  \frac{\phi_{ij}^{\nu_i + \nu_j} }{ \phi_i^{\nu_i} \phi_j^{\nu_j} }  \mathrm{Ga}\biggr(\rho^2| \nu_{ij}, \frac{1}{4\phi_{ij}^2}\biggr). 
\end{align}

Then applying this (scaled) Gamma mixing density to $C_{ij}(\cdot, \cdot; \rho^2)$ yields
\begin{align*}
K_{ij}(\Ell_k, \Ell_m; \phi^2) &= E_{p_0}\biggr\{ \sqrt{p_i(\rho^2| \phi^2) p_j(\rho^2|\phi^2)} /p_0(\rho^2|\phi^2)  C_{ij}(\Ell_k, \Ell_m; \rho^2) | \phi^2 \biggr\}  \\
& = c_{ij}^{\nu}  \frac{\phi_{ij}^{\nu_i + \nu_j} }{ \phi_i^{\nu_i} \phi_j^{\nu_j} } \sigma_i \sigma_j a_{ij} |\bfO_{ij} |^{-1/2}  \\ 
& \quad \times \int \mathrm{Ga}\biggr(\rho^2| \nu_{ij}, \frac{1}{4\phi_{ij}}\biggr) \times \exp\left( - \frac{\bfh_{ij}^\top \bfO_{ij}^{-1} \bfh_{ij} }{\rho^2} \right) \mathrm{d} \rho^2 \\ 
&= \sigma_i \sigma_j a_{ij} c_{ij}^{\nu}   \frac{\phi_{ij}^{\nu_i + \nu_j} }{ \phi_i^{\nu_i} \phi_j^{\nu_j} }  |\bfO_{ij} |^{-1/2} \mathcal{M}(r_{ij}; \nu_{ij}, \phi_{ij}),
\end{align*}
where the last equation uses the scale-mixture trick in Proposition 1.

To derive the  CH cross-covariance, consider $p_i(\rho^2|\phi^2) = \text{Ga}(\nu_i, 1/(4\phi_i^2))$. Observe that 
\begin{align*}
\sqrt{\pi_i(\phi^2) \pi_j(\phi^2)} = c_{ij}^{\alpha, \beta} \mathrm{IG}(\phi^2| \alpha_{ij}, \beta_{ij}^2/4). 
\end{align*} 

It is immediate that 
\begin{align*}
\tilde{K}_{ij}(\Ell_k, \Ell_m) &=  \int \sqrt{\pi_i(\phi^2) \pi_j(\phi^2)}   K_{ij}(\Ell_k, \Ell_m; \phi^2) \mathrm{d}\phi^2 \\
&= \sigma_i \sigma_j a_{ij} c_{ij}^{\nu} |\bfO_{ij}|^{-1/2} \int \mathcal{M}(r_{ij}; \nu_{ij}, \phi) c_{ij}^{\alpha,\beta} \mathrm{IG}(\phi^2| \alpha_{ij}, \beta_{ij}^2/4) \mathrm{d} \phi^2 \\
&= \sigma_i \sigma_j a_{ij} c_{ij}^{\nu} c_{ij}^{\alpha,\beta} |\bfO_{ij}|^{-1/2} \mathcal{CH}(r_{ij}; \nu_{ij},\alpha_{ij},\beta_{ij}).
\end{align*} 
\end{proof}

\subsection{Proof of Theorem~\ref{thm: GLag framework}} \label{app: GLag framework}
\begin{proof}

Since $\varphi(\cdot)$ is an isotropic correlation for any dimension, it follows from Schoenberg's representation \citetext{\citealt{Schoenberg1938}, \citealt[][pp.~44-45]{Stein1999}} that there is a probability measure $\mu$ on $[0,\infty)$ such that 
\begin{align*}
\varphi(r) = \int_0^{\infty} e^{-sr^2} \mu(\mathrm{d}s). 
\end{align*}
To show $\psi(\cdot,\cdot)$ is positive definite, it suffices to establish positive definiteness of 
\begin{align*}
K_s(\bfh, u) := | \bfI_{d\times d} + u^2 \bfLambda|^{-a/(2d)} e^{-s r_u^2},
\end{align*}
for every $s\geq 0$, where $r_u^2: = (\bfh - u\bflambda)^\top \{ \bfI_{d\times d} + u^2 \bfLambda\}^{-1} (\bfh - u\bflambda)$.

For $s>0$, observe that for any symmetric positive definite matrix $B$, it is easy to check that 
\begin{align} \label{eqn: Gaussian identity} 
\int_{\mathbb{R}^d} \exp(i \bfh^\top \bfomega - \bfomega^\top B \bfomega) \mathrm{d} \bfomega = \pi^{d/2} |B|^{-1/2} \exp(-\bfh^\top B^{-1} \bfh/4). 
\end{align}
Then by Eq.~\eqref{eqn: Gaussian identity}, 
\begin{align} \nonumber 
\hat{K}_s(\bfomega, u)&:= \int_{\mathbb{R}^d} e^{-i\bfomega^\top \bfh} K_s(\bfh, u) \mathrm{d} \bfh \\ \nonumber
&= |\bfI_{d\times d}  + u^2 \bfLambda|^{-a/(2d)} \exp(-i u \bfomega^\top \bflambda) \int_{\mathbb{R}^d} \exp\left\{- i \bfomega^\top \bfh - s \bfh^\top (\bfI_{d\times d}  + u^2 \bfLambda)^{-1} \bfh \right\} \mathrm{d} \bfh \\ \nonumber
&= (\pi/s)^{d/2}  |\bfI_{d\times d}  + u^2 \bfLambda|^{1/2-a/(2d)} \exp \left\{ - \frac{\bfomega^\top (\bfI + u^2 \bfLambda) \bfomega }{4s} - i u \bfomega^\top \bflambda \right\} \\ \label{eqn: spatial fourier} 
& = (\pi/s)^{d/2} e^{-\|\bfomega\|^2/ (4s)}  |\bfI_{d\times d}  + u^2 \bfLambda|^{-\frac{a-d}{2d}}  \exp\left\{ - u^2 \bfomega^\top \bfLambda \bfomega /(4s) - i u \bfomega^\top \bflambda \right\}. 
\end{align}

Therefore, by Fourier inversion, 
\begin{align} \label{eqn: fourier representation} 
K_s(\bfh, u) = \frac{1}{(2\pi)^d} \int_{\mathbb{R}^d} e^{i \bfomega^\top \bfh} \hat{K}_s(\bfomega, u) \mathrm{d} \bfomega.
\end{align}
What follows is to establish positive definiteness by definition. 

 For $a>d$, let $k_1, \ldots, k_d$ be the eigenvalues of the positive definite matrix $\bfLambda$. Then $g(u) := | \bfI_{d\times d} + u^2 \bfLambda|^{-(a-d)/(2d)}  = \prod_{i=1}^d (1 + u^2 k_i)^{-(a-d)/(2d)}$, where 
 \begin{align*}
 g_i(u) := (1 + u^2 k_i)^{-(a-d)/(2d)} = \frac{1}{\Gamma((a-d)/(2d))} \int_0^\infty x^{(a-d)/(2d) - 1} \exp \{ - x (1 + u^2 k_i)\} \mathrm{d} x.
 \end{align*}
Each $g_i$ is a positive mixture of Gaussian covariance functions and hence $g(u)$ is positive definite for $a>d$.  When $a=d$, $g(u) = 1$ which is positive definite. 

Note that for every fixed $\bfomega$, the term 
$
\exp\left\{ - u^2 \bfomega^\top \bfLambda \bfomega /(4s)  \right\}
$
is positive definite as a function of $u$. It follows from the Schur product theorem that its product with $g(u)$ is also positive definite. 

For any $n$, let $(\bfh_j, u_j)\in \mathbb{R}^d\times \mathbb{R}$ and coefficients $c_j\in \mathbb{C}$ with $j=1,\ldots, n$, define  $z_j(\bfomega):=c_j \exp[i\bfomega^\top (\bfh_j-u_j \bflambda) ]$. Then it follows from Equations~\eqref{eqn: spatial fourier} and~\eqref{eqn: fourier representation} that  
\begin{align*}
&\sum_{j,k=1}^n c_j \bar{c}_k K_s(\bfh_j-\bfh_k, u_j-u_k) \\
&= \frac{1}{(2\pi)^d} \int_{\mathbb{R}^d}
\sum_{j,k=1}^n c_j \bar{c}_k
 e^{i\bfomega^\top(\bfh_j-\bfh_k)}
 \hat{K}_s(\bfomega,u_j-u_k)\,\mathrm{d}\bfomega \\
&= \frac{1}{(4\pi s)^{d/2}} \int_{\mathbb{R}^d}
 e^{-\|\bfomega\|^2/(4s)}
 \Biggl\{\sum_{j,k=1}^n z_j(\bfomega)\overline{z_k(\bfomega)} \\
&\qquad\qquad\times
 \bigl|\bfI_{d\times d}+(u_j-u_k)^2\bfLambda\bigr|^{-(a-d)/(2d)}
 \exp\left(-\frac{(u_j-u_k)^2\bfomega^\top\bfLambda\bfomega}{4s}\right)
 \Biggr\}\,\mathrm{d}\bfomega \geq 0.
\end{align*}

Therefore $K_s(\bfh, u)$ is positive definite for every $s>0$.

As $s \to 0^+$, $K_s(\bfh, u)$ converges to $K_0(\bfh, u)= | \bfI + u^2 \bfLambda|^{-a/(2d)}$ for every $(\bfh, u)$; thus, $K_0(\bfh, u)$ is also positive definite. The conclusion follows. 

To derive the tail behavior for $\psi$,  for any fixed $\bfh$, as $u\to \infty$, $g(u) \sim |\bfLambda|^{-a/(2d)} |u|^{-a}$. By continuity of  $\varphi$, when $\bflambda \neq \mathbf{0}$, $\varphi(r_u) \sim \varphi((\bflambda^\top \bfLambda^{1/2} \bflambda)^{-1})$ as $u\to \infty$. Combining these two expressions yields the desired result. 

\end{proof}

%\vspace*{-10pt}

%\appendixtwo

\section{Mixture Construction} \label{sec: mixture construction}

\subsection{A Lagrangian Generalized Hyperbolic Model} \label{sec: LagGH}
By specifying a generalized inverse Gaussian (GIG) distribution \citep[e.g.,][]{Barndorff1982, Barndorff1997} for $\rho^2$, we could obtain another class of nonseparable space-time covariance models with closed-form expressions. Let $X$ have a generalized inverse Gaussian distribution, denoted by $X\sim \text{GIG}(p, a, b)$, with density 
\begin{align*}
    p(x) = \frac{(a/b)^{p/2}}{2\mathcal{K}_p(\sqrt{ab})} x^{p-1} \exp\left\{ - \frac{1}{2}(a x + b/x) \right\}, a>0, b>0, p \in \mathbb{R},
\end{align*} 
where $\mathcal{K}_p$ is the modified Bessel function of the second kind. It is easy to see that $p(x)$ has exponential decay as $x$ becomes large. 

Consider the hierarchical model specification 
\begin{gather*} %\label{eqn: LagCH model}
    %\begin{split}
        K(\mathbf{h}, u) = \int C(\mathbf{h}, u; \rho^2) \text{d} G(\rho^2),\quad   \\
        C(\mathbf{h}, u; \rho^2) = E_{\mathbf{v}}[C_s(\mathbf{h}-\mathbf{v}u)], \\
        \mathbf{v} | \rho, \bflambda, \bfLambda \sim \mathcal{N}_d(\bflambda, \rho^2 \bfLambda/2), \quad 
        \rho^2 \sim \text{GIG}(\nu, \phi^2/2, 2\beta^2), 
    %\end{split}
\end{gather*}
where $G$ denotes the probability measure specified by $\rho^2 \sim \text{GIG}(\nu, \phi^2/2, 2\beta^2)$. It can be checked that the resulting nonseparable space-time covariance model has the form 
\begin{align} \nonumber
    K(\mathbf{h}, u) &= \sigma^2  \left|\mathbf{I}_{d\times d} + u^2 \bfLambda \right|^{-1/2} \int \exp \left\{ - \frac{r_u^2 }{\rho^2} \right\} \text{GIG}(\rho^2| p, \phi^2/2, 2\beta^2) \text{d} \rho^2 \\
    &= \sigma^2  \left|\mathbf{I}_{d\times d} + u^2 \bfLambda \right|^{-1/2} \frac{(1 + r_u^2/\beta^2)^{p/2}}{\mathcal{K}_p( \phi \beta)} \mathcal{K}_p\bigr(\phi ( r_u^2 + \beta^2)^{1/2}\bigr) \label{eqn: LagGH}
\end{align}
where $\sigma^2>0, \phi>0, \beta>0, p \in \mathbb{R}$. We call this a \textit{Lagrangian generalized hyperbolic} (L-GH) model since it includes the moment generating function of the generalized hyperbolic probability density function \citep{Barndorff1982}. 

This general model includes several previous space-time covariance models as special cases. For example, it can be easily checked that the model~\eqref{eqn: LagGH} reduces to the Lagrangian Mat\'ern model~\eqref{eqn: LagMatern cov} (with reparameterization) when $p=\nu>0, \phi=1/\beta$. As the modified Bessel function $\mathcal{K}_{\nu}(x)$ has exponential decay as $x$ becomes large, this model also has exponential decay in the tail as $r_u$ becomes large. Thus, this model is no more flexible than the Lagrangian Mat\'ern model~\eqref{eqn: LagMatern cov} for modeling space-time processes in terms of the smoothness behavior and tail decay. The model~\eqref{eqn: LagGH} is also computationally more expensive for statistical inference because it requires estimating one more parameter than the Lagrangian Mat\'ern model~\eqref{eqn: LagMatern cov}.

%%%%%%%%%%%%%

%%%%%%%%%%%%

\subsection{Existing Covariance Function Models} \label{app: existing models} 
\begin{example}[The Lagrangian framework]
\cite{Cox1988} propose the stationary space-time covariance model of  the form 
\begin{align*}%\label{eqn: LagFrame}
C(\mathbf{h}, u)=E_{\mathbf{v}}[\varphi(\|\mathbf{h} - \mathbf{v}u\|)],
\end{align*}
where $\varphi(\cdot)$ is an isotropic covariance function in $\mathbb{R}^d$. This Lagrangian model is a special case of the hierarchical model representation in the sense that the mixing measure in Equation~\eqref{eqn: scale mixture} is a Dirac measure. In general, no explicit formula is available. However, \cite{Schlather2010} has derived a closed-form expression for the Lagrangian model~\eqref{eqn: Lagrangian framework} for $\mathbf{v}| \bfxi=\{\bfmu_v, \bfD_v\} \sim \mathcal{N}_d(\bfmu_v, \bfD_v/2)$ with the following form: 
\begin{align} \label{eqn: LagGauss Cov}
C(\mathbf{h}, u)=\sigma^2  \Bigl|\mathbf{I}_{d\times d} + \frac{u^2}{\rho^2}\bfD_v\Bigr|^{-1/2}
   \exp\!\Biggl\{-\biggl[
      \bigl(\mathbf{h} - u\bfmu_v \bigr)^\top
       \bigl(u^2 \bfD_v + \rho^2 \mathbf{I}_d\bigr)^{-1}
       \bigl(\mathbf{h} - u\bfmu_v\bigr) \biggr] \Biggr\}.
\end{align} 
\end{example}
\begin{remark}
This covariance model has been used as a building block to construct more flexible covariance models \citep{Alegria2017, Porcu2020, Salvana2021, Salvana2023}. However, this covariance function is infinitely differentiable near the origin, which is undesirable for modeling space-time processes \citep{Stein2005}. This limitation could be addressed via hierarchical mixture modeling with the proposed closed-form covariance models in Section~\ref{sec: stcov}. 
\end{remark}

\begin{example}[Mixture models]
\cite{Fonseca2011} consider a special case of the model~\eqref{eqn: Lagrangian framework} with the following space-time covariance function
\begin{align*}
    C(\bfh, u) = \sigma^2 \{ 1 + \|\bfh/a\|^{\alpha} + |u/b|^{\beta} \}^{-\lambda_0} \mathcal{M}\bigl(2\|\bfh\|^{\alpha/2}; \nu, a^{\alpha/2}\bigr) \{ 1 + |u/b|^{\beta}\}^{-\lambda_2}
\end{align*}
where $\sigma^2>0, a>0, b>0, 0<\alpha \beta\leq 2, \lambda_0>0, \lambda_2>0$. $\mathcal{M}$ denotes the isotropic Mat\'ern correlation function in Example~\ref{ex: Matern} of this paper. 
This covariance function is a special case in~\eqref{eqn: Lagrangian framework} after reparameterization with $\bfv = (\rho/\phi) \mathbf{e}$ and $\boldsymbol{\bfe}^\top \mathbf{h}=0$, where  $\bfe^\top{\bfe} \leq 1$ and $\rho>0, \phi>0$. This covariance function is nonseparable but fully symmetric. For the spatial margin, it behaves like the Mat\'ern covariance with flexible origin behavior;  for the temporal margin, it behaves like the Cauchy covariance that is either non-differentiable or infinitely differentiable near the origin. This covariance function can also be generated using our hierarchical modeling approach. In particular, let $C_1, C_2$ be power-exponential functions in space and in time, respectively, each of which can be written as a scale mixture of Gaussian covariance functions with positive stable distributions \citep{Gneiting1997}. The mixing measures for $\rho, \phi$ are $1/\rho^2 \sim \text{Ga}(\lambda_i, a_i)$ and $1/\phi^2\sim \text{IG}(\nu, 1)$. The scale mixture approach with a more complex specification is also used by \cite{FONSECA2011biometrika} to construct a nonstationary covariance function. 
%\end{remark}
\end{example}

\subsection{Proof of Example~\ref{ex: Gneiting}} \label{app: Gneiting}
\begin{proof}
%Note that for any completely monotone function $\varphi(t),t>0$, it has the Bernstein representation \citep[][p.~439]{Feller1966}:
%$$
%\varphi(t)=\int_0^\infty \exp(-rt) \text{d}F(r)
%$$
%for some non-decreasing function $F$.

% Consider the covariance function~\eqref{eqn: Gneiting cov}: 
%  \begin{align*}
%      C(\mathbf{h}, u) = \frac{\sigma^2}{\psi(|u|^2)^{d/2}} \varphi \left( \frac{\|\mathbf{h}\|^2}{\psi(|u|^2)} \right).
%  \end{align*}
Define
\begin{align*}
    C_{\bfomega}(u) &:= \int e^{-i\bfh^\top \bfomega} C(\bfh, u) \text{d} \bfh \\
    &= \sigma^2 \pi^{d/2}  \int_{0}^{\infty} 
    \exp\left\{ - \frac{\|\bfomega\|^2}{4r} \psi(|u|^2)\right\} r^{-d/2} \text{d} F(r) 
\end{align*}
and $\beta:= \psi(|u|^2)/(4r)$. Then $C_{\bfomega}$ has a Gaussian scale mixture form with respect to $r$. 

First observe that 
\begin{align*}
    \int_{\mathbb{R}^d} \exp\left( i\bfh^\top \bfomega - \beta \|\bfomega\|^2 \right) \text{d} \bfomega = \pi^{d/2} \beta^{-d/2} \exp\biggr( - \frac{\|\bfh\|^2}{4\beta} \biggr).
\end{align*}
It then follows from direct computation that 
\begin{align*}
     & \frac{1}{(2\pi)^d} \int_{\mathbb{R}^d} e^{i\bfh^\top \bfomega} C_{\bfomega}(u) \text{d} \bfomega  \\
    &=\sigma^2 (4\pi)^{-d/2}  \int_0^\infty r^{-d/2} 
    \left[ \int_{\mathbb{R}^d} \exp\left( i\bfh^\top \bfomega - \beta \|\bfomega\|^2 \right) \text{d} \bfomega \right] \text{d} F(r) \\
    & = \sigma^2 (4\pi)^{-d/2}  \int_0^\infty r^{-d/2} 
    \left[ \pi^{d/2} \beta^{-d/2} \exp\biggr( - \frac{\|\bfh\|^2}{4\beta} \biggr) \right] \text{d} F(r) \\
    & = \sigma^2  \int_0^{\infty} \psi(|u|^2)^{-d/2} \exp\biggl\{ - \frac{\|\bfh\|^2}{\psi(|u|^2)}r \biggr\} \text{d} F(r) \\
    & = \sigma^2  \psi(|u|^2)^{-d/2} \varphi \left( \frac{\|\mathbf{h}\|^2}{\psi(|u|^2)} \right). 
\end{align*}
Thus 
\begin{align*}
    C(\bfh, u) = \sigma^2 \int_0^{\infty} \psi(|u|^2)^{-d/2} \exp\biggl\{ - \frac{\|\bfh\|^2}{\psi(|u|^2)}r \biggr\} \text{d} F(r), 
\end{align*}
which admits a Gaussian scale mixture form with respect to $r$ under a finite positive measure $F$. This can be further viewed as an instance of the hierarchical mixture formulation in Equations~\eqref{eqn: scale mixture},~\eqref{eqn: Lagrangian framework},~\eqref{eqn: velocity}, where $p(\bfv)$ is taken as a point mass at zero and $\rho^2 = \psi(|u|^2)/r$ in $C_s$. 
\end{proof}

\subsection{Proof of Example~\ref{ex: stein}} \label{app: stein} 
\begin{proof}
Since this covariance falls into the frozen field Lagrangian framework, we take the distribution of $\bfv$ to be a point mass at $\epsilon \bfe$, where $\epsilon>0$ and $\bfe\in \mathbb{R}^d$ is a unit vector. Consider 
\begin{align*}
C_s(\bfh; w) &:=\sigma^2 \exp\biggr(-\frac{\|{\bfh} \|^2}{2w}  \biggr) \\
g(w; u, \nu, \zeta, \phi) &:= \frac{1}{(2\phi^2)^{(\nu + \zeta |u|)} \Gamma(\nu+\zeta |u| ) } w^{(\nu + \zeta |u|)-1} \exp\biggr(-\frac{1}{2\phi^2} w\biggr), w>0. 
\end{align*} 
Let $W\sim \mathrm{Ga}(\nu+\zeta |u| , 1/(2\phi^2))$. 
Using the following properties of the modified Bessel function: 
\begin{align*}
\mathcal{K}_{\nu} (z) = \mathcal{K}_{-\nu}(z), \quad
\mathcal{K}_{\nu}(z) = \frac{z^{\nu}}{2^{\nu+1}} \int_0^{\infty} t^{-\nu -1} \exp\biggr(- t - \frac{z^2}{4t} \biggr) \mathrm{d} t, \quad \nu\in\mathbb{R},\ z>0,
\end{align*}
it is easy to derive that 
\begin{align*}
&\int_0^{\infty} C_s(\bfh - \epsilon \bfe u; w) g(w; u, \nu, \zeta, \phi)  \mathrm{d} w \\
&= \frac{\sigma^2}{(2\phi^2)^{(\nu + \zeta |u|)} \Gamma(\nu+\zeta |u| ) }  \int_0^{\infty} \exp\biggr(-\frac{\|\bfh - \epsilon \bfe u \|^2}{2w}  \biggr)  w^{(\nu + \zeta |u|)-1} \exp\biggr(-\frac{1}{2\phi^2} w\biggr)  \mathrm{d} w \\
& = \frac{\sigma^2}{(2\phi^2)^{(\nu + \zeta |u|)} \Gamma(\nu+\zeta |u| ) }  \int_0^{\infty}  w^{(\nu+\zeta |u|)-1} \exp\biggr(-\frac{\|\bfh - \epsilon \bfe u \|^2}{2w} - \frac{1}{2\phi^2} w \biggr)  \mathrm{d} w \\
& \overset{w = 2\phi^2 t}{=\joinrel=\joinrel=\joinrel=} \frac{\sigma^2}{ \Gamma(\nu+\zeta |u|) } \int_0^{\infty} t^{\nu+\zeta|u| - 1} \exp\biggr\{ - t - \frac{\|\bfh-\epsilon \bfe u\|^2}{4 \phi^2 t} \biggr\} \mathrm{d} t \\
& =  \frac{\sigma^2}{ \Gamma(\nu+\zeta |u|) }  2^{1- (\nu+\zeta|u|)} \left(\frac{\|\bfh - \epsilon \bfe u\|}{\phi} \right)^{\nu+\zeta|u|} \mathcal{K}_{-(\nu+\zeta|u|)}  \left(\frac{\|\bfh - \epsilon \bfe u \|}{\phi} \right) \\
& = \sigma^2 \mathcal{M}(\|\bfh - \epsilon \bfe u \|; \nu + \zeta|u|, \phi).
\end{align*}

Let $B \sim \text{Beta}(\nu, d/2)$, independent of $W$. Direct computation yields that  
$$
E[B^{\zeta |u|}] = \frac{1}{B(\nu, d/2)} \int_0^1 x^{\nu+\zeta|u|-1} (1-x)^{d/2-1} \mathrm{d} x = 
 \frac{\Gamma(\nu+\zeta |u| ) \Gamma(\nu+d/2) }{\Gamma(\nu) \Gamma(\nu+\zeta |u| + d/2) },
$$

Thus, 
\begin{align*}
C(\bfh, u) &= \sigma^2  \frac{\Gamma(\nu+\zeta |u| ) \Gamma(\nu+d/2) }{\Gamma(\nu) \Gamma(\nu+\zeta |u| + d/2) } \mathcal{M}(\|\bfh - \epsilon \bfe u \|; \nu + \zeta|u|, \phi) \\
&= E_B\left\{ B^{\zeta |u|} E_W \left[ \sigma^2  \exp\biggr(-\frac{\|{\bfh} - \epsilon u \bfe \|^2}{2W} \biggr)  \right] \right\},
\end{align*}
as desired. 
\end{proof}
\vspace*{-10pt}
%\appendixthree
\section{Numerical Results} \label{app: numerical results}

\subsection{Nonstationarity Illustration} \label{app: nonstationarity} 

\begin{figure}[!h]
\begin{center}
\makebox[\textwidth][c]{\includegraphics[width=1\textwidth, height=0.45\textheight]{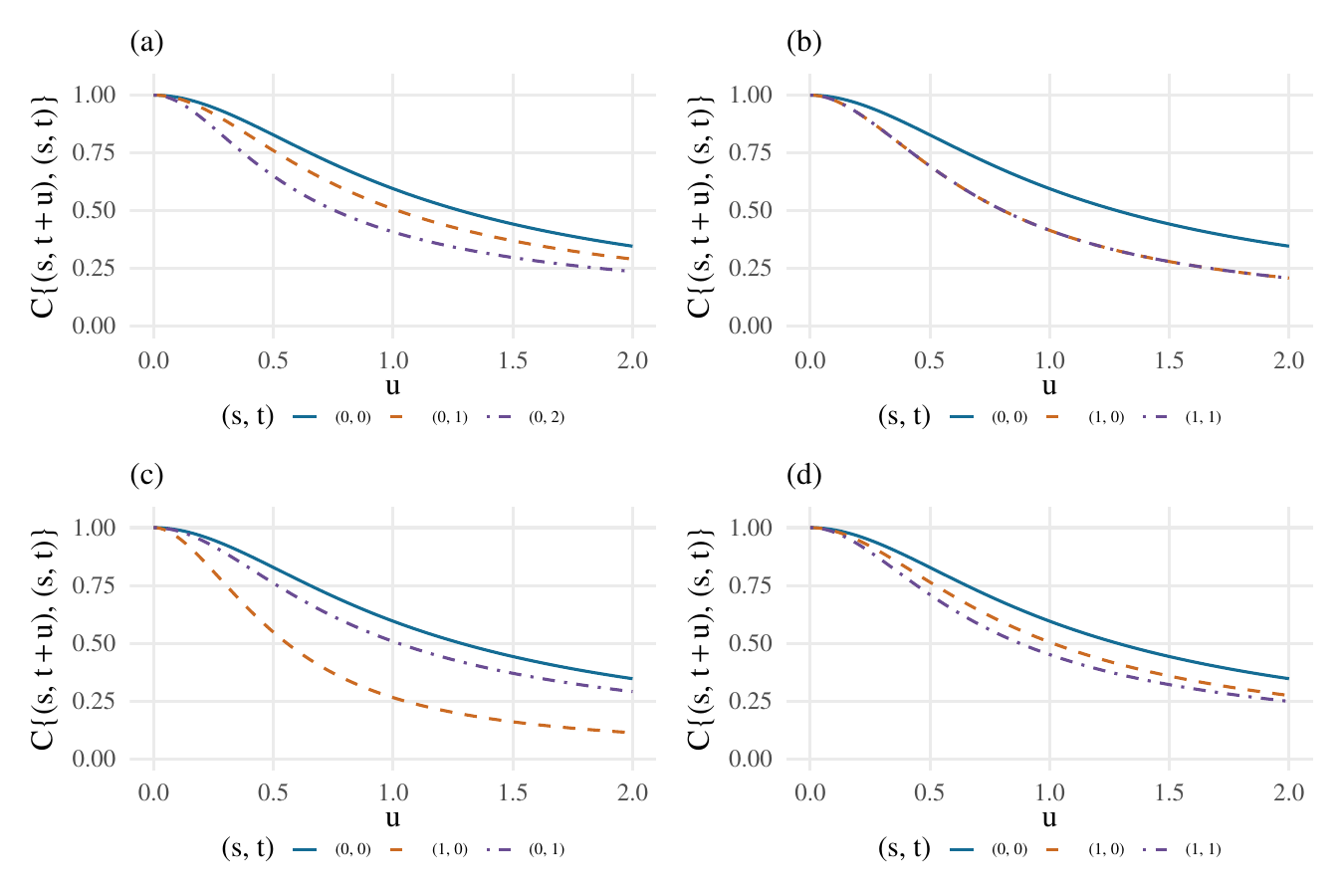}}
\caption{Plot of nonstationary covariance functions. For all panels, the velocity correlation function $\Lambda$ is specified with a stationary Mat\'ern model. Panel (a): $\lambda(s,t)=1, \Lambda(h,u) = \mathcal{M}(\sqrt{h^2+u^2}; 1.5, 1)$; panel (b): $\lambda(s,t)=1+s^2, \Lambda(h,u) = \mathcal{M}(h; 1.5, 1)$; panel (c): $\lambda(s,t)=1+2s, \Lambda(h,u) = \mathcal{M}(\sqrt{h^2+u^2}; 1.5, 1)$; panel (d): $\lambda(s,t)=1 + s^2/2, \Lambda(h,u) = \mathcal{M}(\sqrt{h^2+u^2}; 1.5, 1)$.}
\label{fig: nonstat covariance}
\end{center}
\end{figure}

In Figure~\ref{fig: nonstat covariance}, the velocity field is conditionally specified as a  Gaussian process $v(\cdot) | \rho^2 \sim \mathcal{GP}(\lambda(\cdot), \frac{\rho^2}{2} \Lambda(\cdot))$. $C$ is specified as a Lagrangian Mat\'ern covariance function with smoothness parameter $\nu=1.5$ and range $\phi=1$. The resulting covariance function has the following form
\begin{align*}
C\{(s+h, t+u), (s,t)\} = O^{-1/2} (1 + r )e^{-r}, 
\end{align*}
where 
\begin{gather*}
r = |h-m| / \sqrt{O}, \\
m = (t+u) \lambda(s+h, t+u) - t \lambda(s,t), \\
O =1 + u^2 + 2t(t+u)\{1 - \Lambda(h,u)\}.
\end{gather*}

In panel (a), the resulting space-time covariance $C$ is  stationary in space but nonstationary in time with each spatial location sharing the same temporal covariance structure; in panel (b),   $C$ is nonstationary in space and time; however, the temporal covariance at each spatial location is stationary but different across spatial locations; in panel (c),  $C$ is nonstationary in space and time; however, the spatial covariance at each time is stationary; in panel (d), $C$ is nonstationary in space and time; however, the spatial covariance varies with spatial location while the temporal covariance varies across both spatial locations and time points. In all those panels, the spatial covariance at $t=0$ is stationary, corresponding to the initial field structures with the stationary Gaussian spatial covariance.   

\begin{figure}[!h]
\begin{center}
\makebox[\textwidth][c]{\includegraphics[width=1\textwidth, height=0.28\textheight]{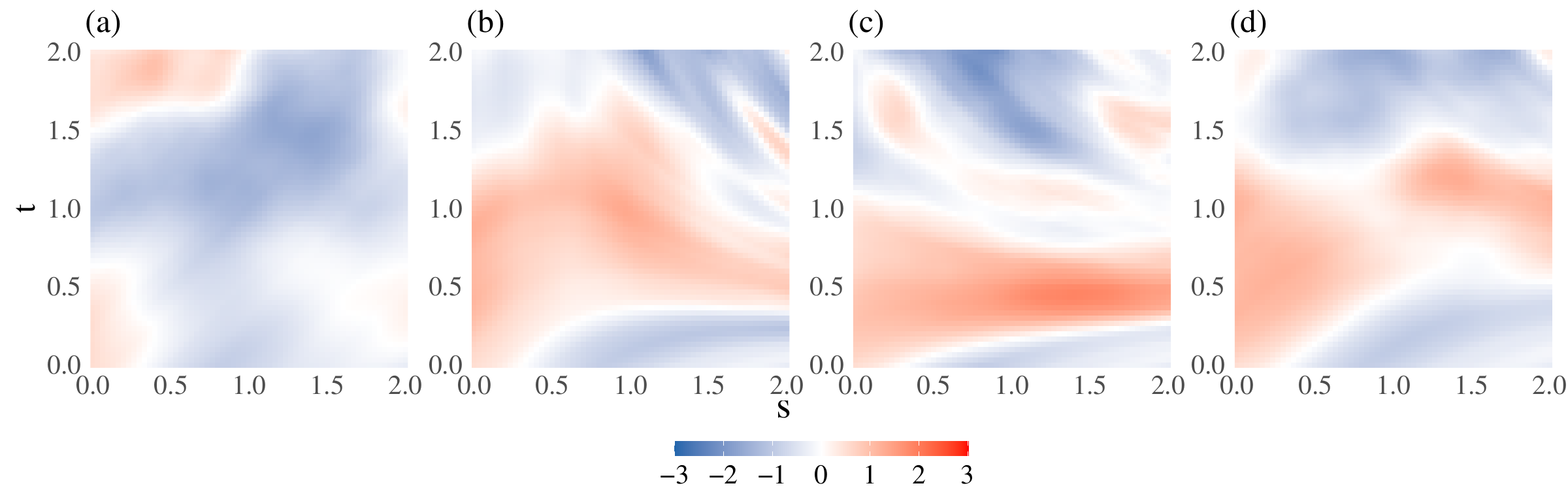}}
\caption{Plot of nonstationary space-time process realizations under the nonstationary Lagrangian Mat\'ern models with $\nu=1.5, \phi=1$. The panels correspond to each of  the velocity models in Figure~\ref{fig: nonstat covariance}, respectively. }
\label{fig: nonstat realization}
\end{center}
\end{figure}

In Figure~\ref{fig: nonstat realization}, the process is simulated with zero mean and a Lagrangian Mat\'ern covariance with smoothness 1.5 and range parameter 1. Each realization corresponds to one of the velocity models in Figure~\ref{fig: nonstat covariance}.  

\newpage

\subsection{Multivariate Illustration} \label{app: multivariate}

\begin{figure}[!htbp]
\begin{center}
\makebox[\textwidth][c]{\includegraphics[width=1\textwidth, height=0.45\textheight]{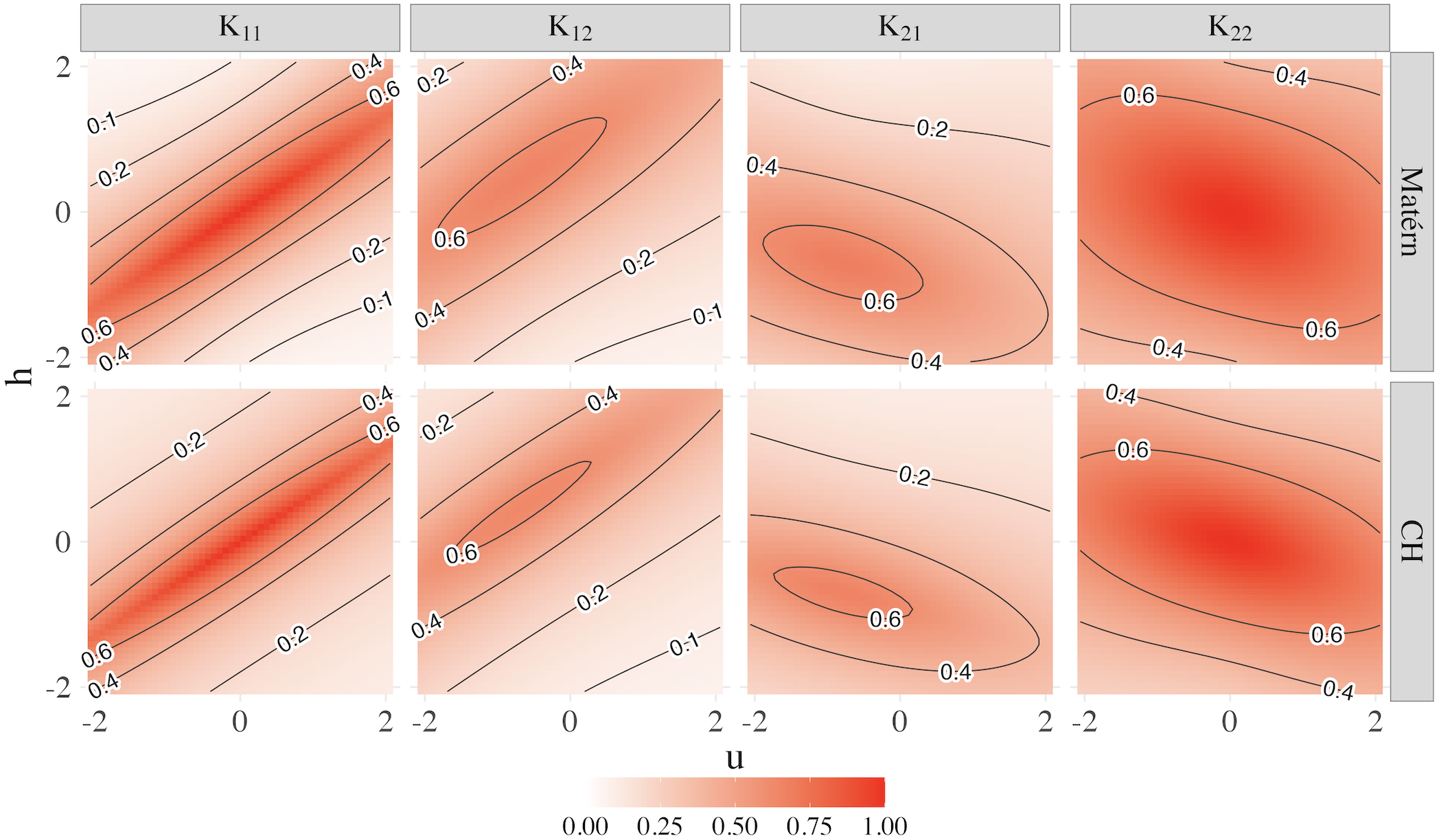}}
\caption{Contour plots of bivariate Lagrangian models with different component velocities $\bfV|\rho^2 \sim \mathcal{N}_2(\bfmu_{\lambda}, \rho^2 \bfSigma_{\lambda}/2)$. Asymmetry between space and time is due to velocities. $t_m=1$ and $t_k=1+u$.}
\label{fig: mvn multiple velocity}
\end{center}
\end{figure}

Figure~\ref{fig: mvn multiple velocity} shows the asymmetric covariance structures between space and time due to different velocity components. The parameter setting is $\sigma_1=\sigma_2=1, \nu_1=0.6, \nu_2=1.8, \phi_1=0.85, \phi_2=0.9, \alpha_1=0.65, \alpha_2=1.5, \beta_1=0.85, \beta_2=1.5, a_{12}=a_{21}=0.98, (\bflambda_1, \bflambda_2)=(0.65, -0.25), \Lambda_{11}=0.18, \Lambda_{22}=0.28, \Lambda_{12}=\Lambda_{21}=0.06$. 

\begin{figure}[!htbp]
\begin{center}
\makebox[\textwidth][c]{\includegraphics[width=1\textwidth, height=0.45\textheight]{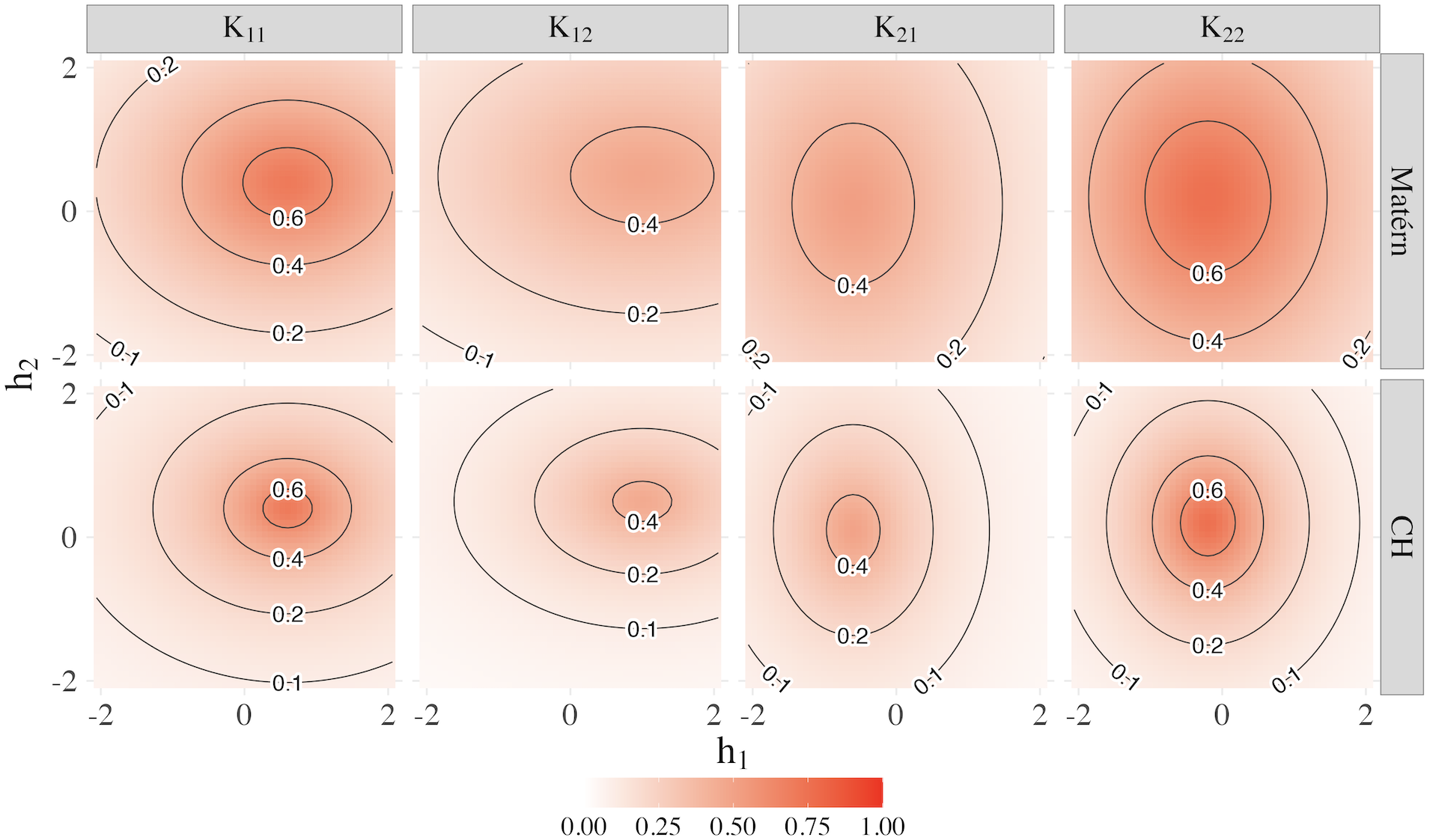}}
\caption{Contour plots of bivariate Lagrangian models at a fixed time lag. Asymmetry between different processes in space is exhibited. $t_m=0.5, t_k=1.5$.}
\label{fig: mvn asymmetry}
\end{center}
\end{figure}

Figure~\ref{fig: mvn asymmetry}  shows the asymmetric structure between different processes at a fixed time lag. The parameter setting is  
$\sigma_1=\sigma_2=1, \nu_1=1, \nu_2=1.5, \phi_1=\phi_2=1, \alpha_1=1, \alpha_2=1.5, \beta_1=\beta_2=1, a_{12}=a_{21}=0.9, \bflambda_1=(0.6, 0.4)^\top, \bflambda_2=(-0.2, 0.2)^\top, \Lambda_{11}=\text{diag}(0.8,0.1), \Lambda_{22}=\text{diag}(0.1,0.6), \Lambda_{12}=\Lambda_{21}=0.1\mathbf{I}_2$.

\vfill
\newpage

\subsection{General Asymmetric Space-Time Models} \label{app: general asymmetric models}

\begin{figure}[!htbp]
\begin{center}
\makebox[\textwidth][c]{\includegraphics[width=1\textwidth, height=0.4\textheight]{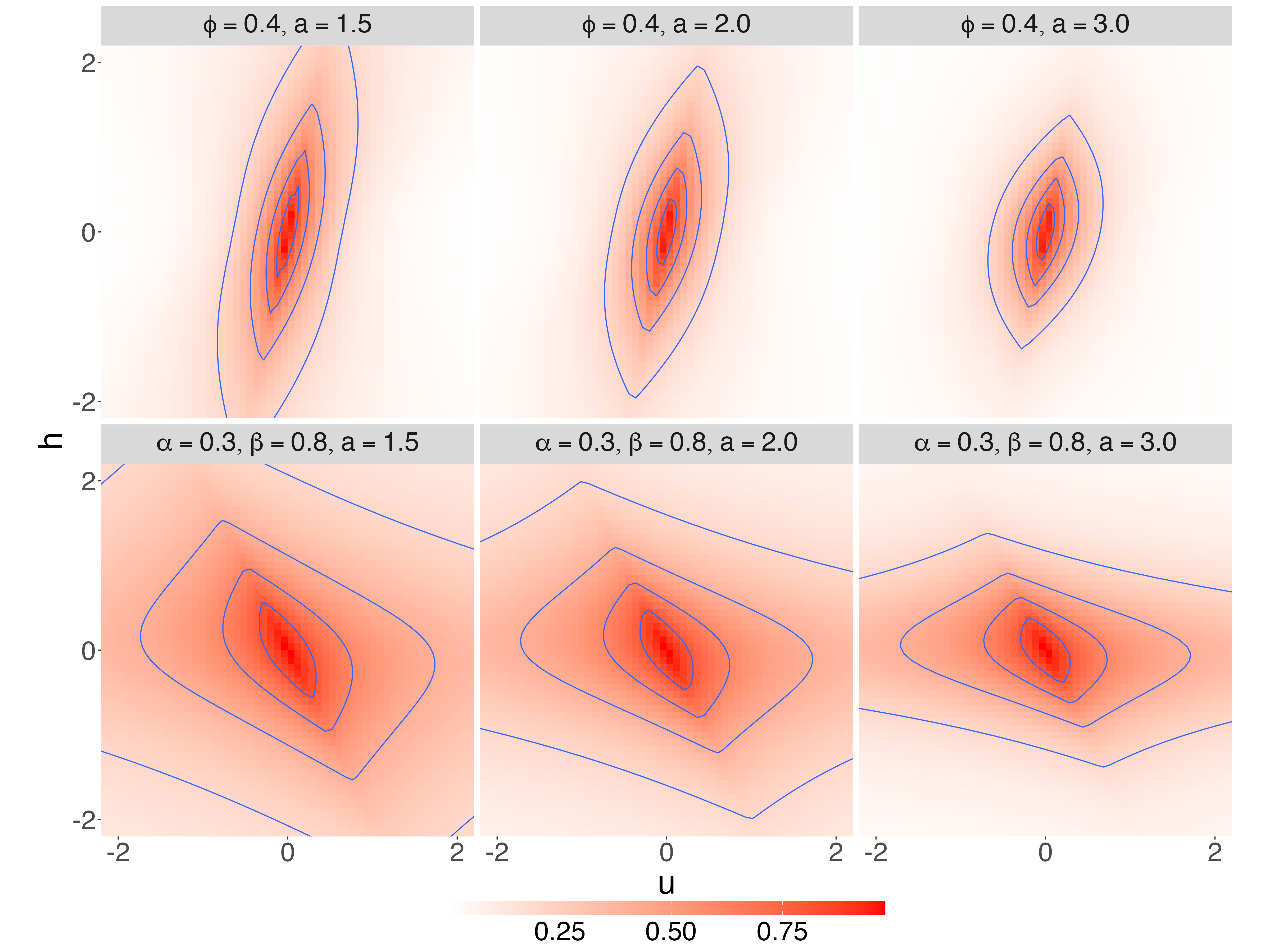}}
\caption{Contour plots of generalized Lagrangian Mat\'ern (first row) and generalized Lagrangian CH models (second row) over the space-time domain $[-2,2]\times [-2, 2]$. All panels share the same parameters $\sigma^2=1, \nu=0.5, \bfLambda = 1$. The generalized Lagrangian Mat\'ern model has asymmetric parameters $\bflambda=0.2$ and the generalized Lagrangian CH model has asymmetric parameters $\bflambda=-0.5$. All other parameter settings are specified in the title of each panel. \vspace{-.6cm}}
\label{fig: GL contours}
\end{center}
\end{figure} %\vspace{-1cm} 
\vfill
\newpage

\subsection{Model Fitting} \label{app: model fitting} 

\begin{table}[!ht]
\centering
\normalsize
   \caption{Model comparison for Irish wind data under different covariance models. AIC represents the Akaike information criterion and BIC represents the Bayesian information criterion. The grouped Vecchia approximation is used for likelihood evaluation with $m=50$ neighbors.}
  {\resizebox{1.0\textwidth}{!}{%
  \setlength{\tabcolsep}{1.0em}
   \begin{tabular}{l c c c c} 
   \toprule \noalign{\vskip 1.5pt} 
Model & log-likelihood & AIC & BIC & \# of cov. parameters \\ \noalign{\vskip 1.5pt} 
\midrule \noalign{\vskip 1.5pt} 
\multicolumn{5}{c}{ \ul{$m=50$}  }\\  \noalign{\vskip 3.5pt} 
Lagrangian Gaussian  & -29,813 & 59,642 & 59,716 & 8\\ \noalign{\vskip 2.5pt} 
Lagrangian Mat\'ern  & -13,115 & 26,249 & 26,331 & 9\\ \noalign{\vskip 2.5pt} 
Lagrangian CH   & -13,114 & 26,249 & 26,341 & 10 \\ \noalign{\vskip 2.5pt} 
\bottomrule
   \end{tabular}%
   }}
   \label{table: irish wind m=50}
\end{table}  

\begin{table}[!ht]
\centering
\normalsize
   \caption{Parameter estimates for Irish wind data under different covariance models. The grouped Vecchia approximation is used for likelihood evaluation with 150 neighbors. Note that numbers less than $10^{-8}$ are reported as 0.0. Identical displayed estimates may differ before rounding.}
  {\resizebox{1.0\textwidth}{!}{%
  \setlength{\tabcolsep}{.5em}
   \begin{tabular}{l l l l l l l l l l l } 
   \toprule \noalign{\vskip 1.5pt} 
Model & $\phi$ or $\beta$ & $\alpha$ & $\nu$ & $\tau^2$ & $\sigma^2$ & $\lambda_0$ & $\theta_0$ & $\theta_1$ & $\lambda_1$ & $\lambda_2$  \\ \noalign{\vskip 1.5pt} 
\midrule \noalign{\vskip 2.5pt} 
Lagrangian Gaussian  & 3.04 & \rule[0.5ex]{.25cm}{0.5pt} & \rule[0.5ex]{.25cm}{0.5pt} & 0.082 & 0.54 & $1.8\times 10^{-7}$ & -2.60 & 0.68 & 0.0 & 6.90 \\ \noalign{\vskip 2.5pt} 
Lagrangian Mat\'ern  & 9.71 & \rule[0.5ex]{.25cm}{0.5pt} & 0.37 & $8.4\times 10^{-6}$ & 0.58 & 0.0 & -3.09 & 0.75 &0.0 & 0.84 \\ \noalign{\vskip 2.5pt} 
Lagrangian CH   & 37.0 & 3.43 & 0.40 & 0.008 & 0.43 & 0.0 & -3.14 & 0.71 & 0.0 & 1.37  \\ \noalign{\vskip 2.5pt} 
\bottomrule
   \end{tabular}%
   }}
   \label{table: irish wind param}
\end{table}

\begin{table}[!ht]
\centering
\normalsize
   \caption{Parameter estimation for air temperature data under different covariance models. The grouped Vecchia approximation is used for likelihood evaluation with 150 neighbors. Note that numbers less than $10^{-8}$ are reported as 0.0. Identical displayed estimates may differ before rounding.}
  {\resizebox{1.0\textwidth}{!}{%
  \setlength{\tabcolsep}{.5em}
   \begin{tabular}{l l l l l l l l l l l } 
   \toprule \noalign{\vskip 1.5pt} 
Model & $\phi$ or $\beta$ & $\alpha$ & $\nu$ & $\tau^2$ & $\sigma^2$ & $\lambda_0$ & $\theta_0$ & $\theta_1$ & $\lambda_1$ & $\lambda_2$  \\ \noalign{\vskip 1.5pt} 
\midrule \noalign{\vskip 2.5pt} 
Lagrangian Gaussian & 0.35  & \rule[0.5ex]{.25cm}{0.5pt}  &  \rule[0.5ex]{.25cm}{0.5pt} & 0.84 & 0.54 & $6.7\times 10^{-7}$ & -3.10 & 0.70 & $6.0\times 10^{-3}$ & $5.1\times 10^{-6}$ \\ \noalign{\vskip 2.5pt}
Lagrangian Mat\'ern  & 4.73 & \rule[0.5ex]{.25cm}{0.5pt} & 0.12 & $2.2\times 10^{-8}$ & 1.22 & $1.0\times 10^{-5}$ & -3.11 & 3.14 &1.04 & $1.1\times 10^{-4}$ \\ \noalign{\vskip 2.5pt} % result from version 0.1.2
Lagrangian CH   &15.8  & 3.12 &0.12  & $2.2\times 10^{-8}$  & 1.22  & $1.0\times 10^{-5}$  & -3.11  & 3.14 & 1.03  &  $1.1\times 10^{-4}$ \\ \noalign{\vskip 2.5pt} 
\bottomrule
   \end{tabular}%
   }}
   \label{table: AT param}
\end{table}  

\newpage 
\subsection{Prediction Results} \label{app: prediction results}

We use the air temperature dataset to compare the predictive performance of the Lagrangian Gaussian, Mat\'ern, and CH models. As the likelihoods are quite similar for the Lagrangian Mat\'ern and CH models, it is expected that there is not much difference in terms of interpolative predictive performance but we may expect better predictive performance in terms of long-range predictions. Thus, we will focus on comparing predictions between Lagrangian Gaussian and Lagrangian Mat\'ern/CH models. Since there are 363 distinct observed days of data, we use the first 80\% of observed days for training, ending on the 290th observed day (i.e., day 292). The resulting number of training data points is 157,968. The testing set contains 3,096 data points on days 293--299, corresponding to forecast horizons of 1--7 days, and 8,757 data points on days 307--322, corresponding to forecast horizons of 15--30 days. These two testing datasets could be used to test predictive performance at different forecast horizons. For the training dataset, we perform maximum likelihood estimation via grouped Vecchia approximation with $m=50$ nearest neighbors for Lagrangian Gaussian, Mat\'ern, and CH models.

The results are summarized in Table~\ref{table: AT prediction}. 
As expected, the Lagrangian Mat\'ern and CH models give much better results than the Lagrangian Gaussian model in terms of RMSE, log-score, and empirical coverage. Because of the strong assumption on smoothness under the Lagrangian Gaussian model, the average length of prediction intervals is much shorter, indicating overconfidence in prediction with lower empirical coverage. Between the Lagrangian Mat\'ern and CH models, there is essentially no difference, which is expected from the likelihood fit since they yield nearly the same likelihood indicating that it suffices to use exponential decay as in Mat\'ern to model the underlying space-time process from the data. An interesting observation is that forecasts at both 1--7-day and 15--30-day horizons have empirical coverage close to the nominal level of 0.95.

% Required packages: booktabs, multirow, graphicx

\begin{table}[!ht]
\centering
\normalsize
\caption{Predictive performance for air temperature data.
RMSE = root mean squared error;
log-score = mean negative log predictive density;
CVG(95\%) = empirical coverage of 95\% prediction intervals;
LPI(95\%) = mean length of 95\% prediction intervals.
The grouped Vecchia approximation is used with $m=50$. 
}
\label{table: AT prediction}

\begingroup
\setlength{\tabcolsep}{8pt}
\renewcommand{\arraystretch}{1.15}
\resizebox{\textwidth}{!}{%
\begin{tabular}{llcccc}
\toprule
Model & Forecast range & RMSE & log-score & CVG(95\%) & LPI(95\%) \\
\midrule
\multirow{2}{*}{Lagrangian Gaussian}
  & 1--7 days   & 1.059 & 1.55 & 0.86 & 3.19 \\
  & 15--30 days & 1.076 & 1.51 & 0.92 & 3.78 \\
\addlinespace[6pt]
\multirow{2}{*}{Lagrangian Mat\'ern}
  & 1--7 days     & 0.998 & 1.40 &0.96 & 4.00 \\
  & 15--30 days & 1.059 & 1.48 & 0.97 & 4.40 \\ 
  \addlinespace[6pt]
\multirow{2}{*}{Lagrangian CH}
  & 1--7 days   & 0.998 & 1.40 & 0.95 & 3.99 \\
  & 15--30 days & 1.058 & 1.47 & 0.97 & 4.39 \\
\bottomrule
\end{tabular}%
}
\endgroup
\end{table}

\end{document}